\documentclass[11pt]{article}

\usepackage[margin=1cm]{caption}
\usepackage[font=small,labelfont=bf]{caption}

\newcommand{\tr}{\,\text{tr}\,}%
\newcommand{\argth}{\,\text{argth}\,}%
\newcommand{\argsh}{\,\text{argsh}\,}%
\usepackage{float} % added by E
\usepackage{authblk}

\usepackage{cite} 
\usepackage{amsmath}
\usepackage{amsfonts}

\usepackage{tikz}

\begin{document}

\title{Inhomogeneous Gaussian Free Field inside the interacting arctic curve}

\author[1,3]{Etienne Granet}

\author[1,4]{Louise Budzynski}

\author[4]{J\'er\^ome Dubail}

\author[1,2,3]{Jesper Lykke Jacobsen}

\affil[1]{Laboratoire de Physique Th\'eorique, D\'epartement de Physique de l'ENS, \protect \\
Ecole Normale Sup\'erieure, Sorbonne Universit\'e, CNRS, \protect \\
PSL Research University, 75005 Paris, France}

\affil[2]{Sorbonne Universit\'e, \'Ecole Normale Sup\'erieure, CNRS, \protect \\
Laboratoire de Physique Th\'eorique (LPT ENS), 75005 Paris, France }

\affil[3]{Institut de Physique Theorique, CEA Saclay, 91191 Gif-sur-Yvette, France}

\affil[4]{Laboratoire de Physique et Chimie Th\'eoriques, CNRS UMR 7019, \protect \\
Universit\'e de Lorraine, F-54506 Vandoeuvre-les-Nancy, France}

\date{\today}
\maketitle

\abstract{The six-vertex model with domain-wall boundary conditions is one representative of a class of two-dimensional lattice statistical mechanics models that exhibit a phase separation known as the {\it arctic curve} phenomenon. In the thermodynamic limit, the degrees of freedom are completely frozen in a region near the boundary, while they are critically fluctuating in a central region. The arctic curve is the phase boundary that separates those two regions. Critical fluctuations inside the arctic curve have been studied extensively, both in physics and in mathematics, in {\it free models} (i.e., models that map to free fermions, or equivalently to determinantal point processes). Here we study those critical fluctuations in the {\it interacting} (i.e., not free, not determinantal) six-vertex model, and provide evidence for the following two claims:
\begin{enumerate}
\item[(i)] the critical fluctuations are given by a Gaussian Free Field (GFF), as in the free case, but
\item[(ii)] contrarily to the free case, the GFF is {\it inhomogeneous}, meaning that its coupling constant $K$ becomes position-dependent, $K \rightarrow K({\rm x})$.
\end{enumerate}
The evidence is mainly based on the numerical solution of appropriate Bethe ansatz equations with
an imaginary extensive twist, and on transfer matrix computations, but the second claim is also
supported by the analytic calculation of $K$ and its first two derivatives in selected points.
Contrarily to the usual GFF, this inhomogeneous GFF is not defined in terms of the Green's function of the Laplacian $\Delta = \nabla \cdot \nabla$ inside the critical domain, but instead, of the Green's function of a generalized Laplacian $\Delta = \nabla \cdot \frac{1}{K} \nabla$ parametrized by the function $K$.
Surprisingly, we also find that there is a change of regime when $\Delta \leq -1/2$, with $K$ becoming singular at one point.

%The evidence is mainly based on the numerical solution of appropriate Bethe ansatz equations with an imaginary extensive twist, and on transfer matrix computations, but the second claim is also supported by the analytic calculation of $K$ and its first two derivatives in selected points. For $\Delta\leq -1/2$ we find that there is a change of regime that causes $K(m_x,m_y)$, seen as a function of the magnetizations, to be singular at the origin. Contrarily to the usual GFF, this inhomogeneous GFF is not defined in terms of the Green's function of the Laplacian $\Delta = \nabla \cdot \nabla$ inside the critical domain, but instead, of the Green's function of a generalized Laplacian $\Delta = \nabla \cdot \frac{1}{K} \nabla$ parametrized by the function $K$.}

\tableofcontents

\section{Introduction}

The six-vertex model with domain-wall boundary conditions \cite{korepin1982calculation,AGIzergin,izergin1992determinant,KorepinZinnJustin,zinn2000six,zinn2002influence,bogoliubov2002boundary,palamarchuk20106,galleas2010functional,cugliandolo2015six,lyberg2018phase,keesman2017numerical,lyberg2017density} is an example of a two-dimensional statistical mechanics problem that exhibits a limit shape phenomenon \cite{vershik1977asymptotics,logan1982variational,rottman1984statistical,nienhuis1984b,kenyon2007limit,colomo2010limit,reshetikhin2016limit}: the emergence, in the thermodynamic limit, of two or more spatially separated phases. For introductions to this vast topic, see for instance the Lecture Notes by Kenyon on limit shape phenomena in dimer models  \cite{kenyon2009lectures} or on the six-vertex model by Reshetikhin \cite{reshetikhin2010lectures}. The phrase {\it arctic circle}, or more generally {\it arctic curve}, was coined in Ref. \cite{jockusch1998random} to name the boundary that separates these regions. The determination of this curve is, in general, a difficult and very rich problem that has attracted a wide attention \cite{ColomoPronko,colomo2010arctic,colomo2011algebraic,colomo2015thermodynamics,ColomoSportiello,reshetikhin2017integrability,di2018arctic,di2018arctic2}. For the six-vertex model, an exact formula for the arctic curve was derived by Colomo and Pronko in Ref. \cite{ColomoPronko}. \\

For concreteness, let us introduce some notations. Throughout the paper, we deal with the {\it isotropic six-vertex model}, defined by the non-vanishing Boltzmann weights shown in Fig. \ref{fig:6v}.
\begin{figure}[H]
 \begin{center}
\includegraphics[scale=0.3]{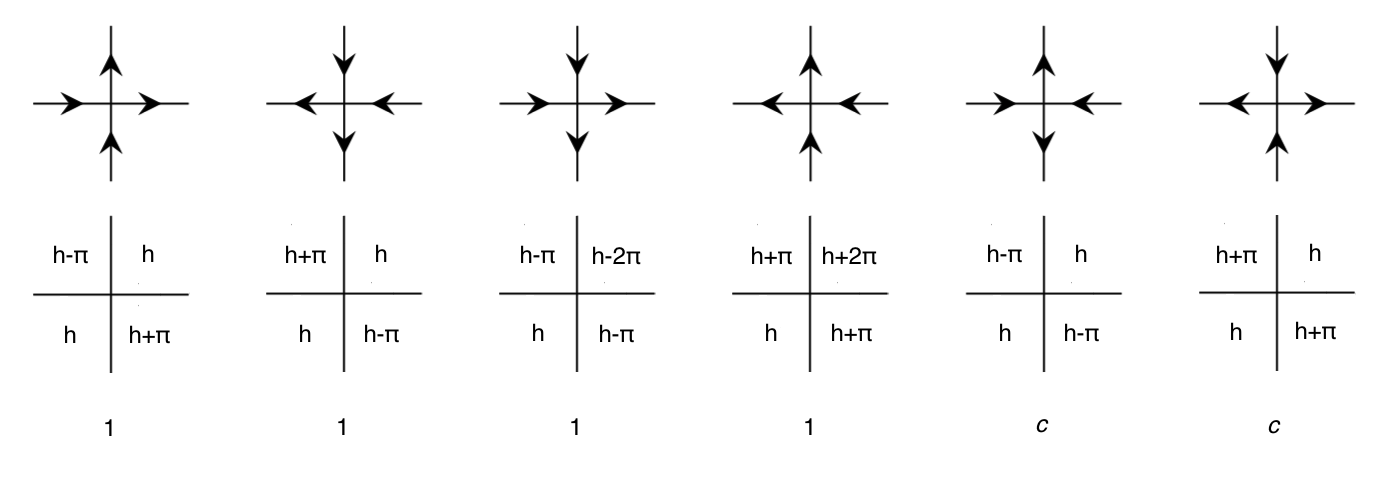}
\end{center}
\caption{The six vertices of the six-vertex model. The top row shows the arrow configurations, while the bottom row shows the corresponding height field configurations. For simplicity, we work with the {\it isotropic six-vertex model}, which has the same Boltzmann weight $1$ for the four first vertices, and another weight $c$ for the last two vertices. This isotropic model is obviously symmetric under exchange of the $x$ and $y$ directions.}
 \label{fig:6v}
\end{figure}
Throughout the paper, the model then depends on a single real positive parameter $c$. To make connections with other references, it is also convenient to use the {\it anisotropy parameter} $\Delta$ defined as
\begin{equation}
	\Delta  = \frac{2 - c^2}{2}  ,
\end{equation}
and it is well known that the model can be mapped on free fermions iff $\Delta = 0$. In what follows, we focus exclusively on the critical regime
\begin{equation}
	-1 <\Delta < 1 .
\end{equation}
For information about the case $|\Delta | \geq 1$, we refer the reader to Ref. \cite{reshetikhin2010lectures}. Let us now quickly review the standard scenario that allows us to understand the phase separation phenomenon in this model.  \\

One considers a square lattice of $N \times N$ vertices, with {\it domain-wall boundary conditions}, see Fig. \ref{fig:DW}. We find that it is convenient to call the lattice spacing $a_0$. We are interested in taking the limit $a_0 \rightarrow 0$, keeping the length of the system $L = N a_0$ fixed. Of course, this is nothing but the thermodynamic limit $N = L/a_0 \rightarrow \infty$, but it is more convenient to think of the lattice as being rescaled such that its global shape remains unchanged. In the limit $a_0$, the lattice points densely fill the square $[0,L]^2$, and we write their position as ${\rm x} = (x,y) \in [0,L]^2$.
\begin{figure}[H]
 \begin{center}
\includegraphics[scale=0.4]{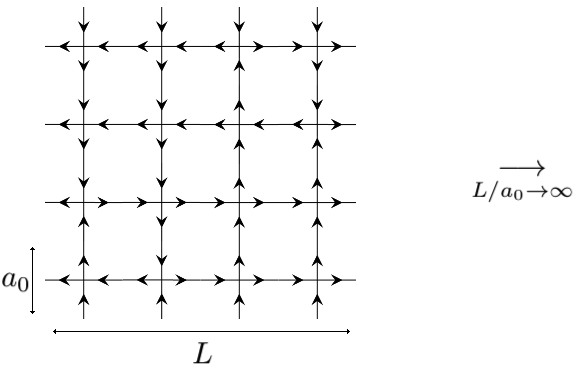}
\quad  \qquad 
\includegraphics[scale=0.21]{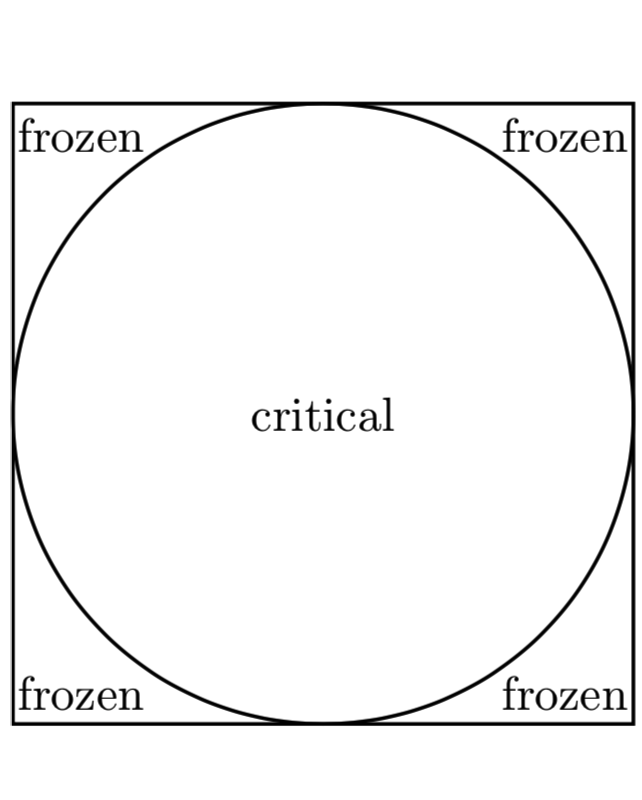}
\end{center}
	\caption{Left: example of a valid configuration of the six-vertex model with domain wall boundary conditions. Right: cartoon of the arctic circle phenomenon. The critical region is a disc only at the free fermion point $\Delta=0$; in general it has a more complicated shape \cite{ColomoPronko}.}
	\label{fig:DW}
\end{figure}

\iffalse
\begin{figure}[h]
	\begin{center}

	\begin{tikzpicture}[transform canvas={scale=1}]
		\draw[thick] (-2.5,-2.5) rectangle (2.5,2.5);
		\draw[thick] (0,0) circle (2.5);
		\draw (0,0) node{critical};
		\draw (-1.95,-2.25) node{frozen};
		\draw (1.95,-2.25) node{frozen};
		\draw (-1.95,2.25) node{frozen};
		\draw (1.95,2.25) node{frozen};
	\end{tikzpicture}
	\end{center}
\end{figure}
\fi
%There are two spatially separated regions, a central one that is critically fluctuating, and another around the boundary where the degrees of freedom are frozen.
%This problem has a long history in physics, mathematics and computer science. The phenomenon can be understood as follows.

For a lattice mesh $a_0 \sim 1/N$ sufficiently small, one can adopt a coarse-grained description, where lattice configurations of the six-vertex model are replaced by a continuous field ${\rm m} = (m_x, m_y)$, where $m_x$ and $m_y$ represent the locally averaged magnetization around a given point, along the horizontal and vertical axis respectively. The magnetization $m_x$ (resp. $m_y$) is defined as $+1$ for a up arrow (resp. right arrow), and as $-1$ for a down arrow (resp. left arrow). The ice rule, or Kirchhoff's law, that is satisfied by the six-vertex model at the microscopic level, translates to the following conservation law on those components, 
\begin{equation}
	\label{eq:kirchhoff}
	  \partial_x m_y +  \partial_y m_x  = 0.
\end{equation}
The constraint is conveniently solved by introducing a {\it height field} ${\rm x} \mapsto h({\rm x})$, defined for ${\rm x}$ in $[0,L]^2$, such that
\begin{equation}
	\label{eq:height}
	m_x = \frac{1}{\pi}\partial_x h \qquad m_y =  -  \frac{1}{\pi} \partial_y h .
\end{equation}
Notice that this defines the height field only up to a global additive constant, which will later be fixed by the boundary conditions. Once this global additive constant is fixed, there is a one-to-one correspondance between the height configurations and arrow configurations. The factor $1/\pi$ is included in order to match the conventions of Ref. \cite{brun2017inhomogeneous}.

The conservation law (\ref{eq:kirchhoff}) can be checked by direct inspection of the height configurations in the second line of Figure~\ref{fig:6v}.
Note that the two terms $\partial_x m_y$ and $\partial_y m_x$ vanish individually for the first four vertices, whereas for the last two only their sum vanishes.

Here we are defining the continuous height field $h({\rm x})$ directly out of the coarse-grained magnetization ${\rm m} ({\rm x})$, but a discrete height field can also be defined on the lattice, see Fig. \ref{fig:6v}. The discrete height field lives on the faces of the model, and is defined such that $h$ jumps by $\pm \pi$ across each lattice edge, depending on its orientation. The continuous height field is then recovered by locally averaging the discrete one around a given vertex. Notice that, because of the underlying lattice model, neither of the two components of the gradient of $h({\rm x})$ can exceed $\pi /a_0$ in absolute value. So, in this coarse-grained picture, the six-vertex model is described by a continuous height field subject to a constraint,
\begin{equation}
	{\rm x} \in [0,L]^2 \longmapsto h({\rm x}) \in \mathbb{R}  \qquad {\rm s.t. } \quad  (\partial_x h, \partial_y h)  \in \left[-\frac{\pi}{a_0},\frac{\pi}{a_0}\right]^2 .
\end{equation}

The {\it arctic curve} phenomenon can then be understood as follows. In terms of the height field, the domain-wall boundary conditions read
\begin{equation}
	\label{eq:bc}
	\left\{ \begin{array}{rcl}
		h (x , 0) &=& \pi x/a_0\\
		h (x, L)&=& \pi (L-x)/a_0 \\
		h (0, y)&=& \pi y/a_0 \\
		h (L, y)&=& \pi (L-y)/a_0 .
	\end{array} \right.
\end{equation}
Notice that the gradient of $h$ is maximal along the boundary. The average height field is then obtained by minimizing the free energy functional
\begin{equation}
	\label{eq:functional}
	\underset{{h}}{\rm min} \,  \int_{[0,L]^2} dx dy  \, F\left(  \frac{1}{\pi} \partial_x h, -\frac{1}{\pi} \partial_y h \right)\,,
\end{equation}
where $h$ satisfies the above boundary conditions and the constraint. More details on the free energy $F(m_x, m_y)$ (which is also called {\it surface tension} in some references \cite{kenyon2009lectures}) will be given in the following sections; for now, let us simply stress that it has been studied in earlier references in the interacting six-vertex model, see e.g. Refs. \cite{noh1996finite,reshetikhin2010lectures}, but is not known analytically outside $\Delta=0$.

It turns out that the solution to this minimization problem is a height configuration ${\rm x} \mapsto h({\rm x})$ that has maximal slope in a finite region around the square's boundary. That region is the one where the degrees of freedom are frozen. The other region, which we call $\Omega$, is the one where the slope is {\it not} maximal, implying that the height field can fluctuate around its mean value. The curve that separates the two regions is the {\it arctic curve}, which was calculated by Colomo and Pronko for the six-vertex model \cite{ColomoPronko}, see also Ref. \cite{ColomoSportiello} and the other references cited above. \\

Now let us turn to the {\bf objective of this paper}. We want to describe the fluctuations of the height field inside the domain $\Omega$. These fluctuations have been studied in detail in the free fermion case in Refs. \cite{petrov2015asymptotics,RKenyon,2015arXiv151008248D,bufetov2016fluctuations,gorin2017bulk}, with the conclusion that the fluctuations are always given by a Gaussian Free Field (GFF) \cite{sheffield2007gaussian} in a certain metric (we will discuss the metric shortly). This problem was revisited recently by some of us and our collaborators \cite{AllegraDubailStephanViti}, also for free fermions, motivated by connections with out-of-equilibrium problems in one-dimensional quantum systems \cite{dubail2017conformal,DubailStephanCalabrese}. The GFF also describes models of interacting dimers on a periodic lattice \cite{alet1,alet2,Toninelli1,Toninelli2,Toninelli3}, thus without the arctic curve phenomenon, but with a uniform coupling constant that depends on the interaction. To our knowledge, not much is known about fluctuations in the critical region inside the arctic curve in interacting models, and it is the purpose of this paper to partially fill this gap. The study is mainly numerical, but presents also some analytical parts.

There are two basic claims that we want to check numerically. Both are a consequence of the following argument (which some of us used also in closely related works  \cite{DubailStephanCalabrese, brun2017inhomogeneous}): assuming that one knows the mean height configuration $h_{0}$, that is the one that minimizes the total free energy, one can expand to second order around that minimum\footnote{There is a subtlety about the definition of the minimum for finite $L/a_0$ which we overlook here, and which we defer to appendix \ref{app:propermin}.}, and that naturally gives a Gaussian action that should capture the fluctuations in the critical region. To elaborate, a small fluctuation around the mean value
$$
h({\rm x}) = h_{0} ({\rm x})  +   \delta h ({\rm x})
$$
in the critical region $\Omega$ will come with an excess of total free energy given by
\begin{equation}
\label{eq:action0}
 \frac{1}{2} \int_\Omega dx dy \left( \begin{array}{cc} \partial_x \delta h  & \partial_y \delta h  \end{array}   \right)   ({\rm Hess} \, F)   \left( \begin{array}{c} \partial_x \delta h  \\ \partial_y \delta h  \end{array}   \right) \, + \,  \dots, 
\end{equation}
where the '$\dots$' stand for higher order terms that are all irrelevant in the renormalization group sense. We therefore expect that these terms can be dropped in the thermodynamic limit. ${\rm Hess} \, F$ is the hessian of the free energy $F(m_x, m_y)$, which we define as follows for convenience (see also Eqs. (\ref{eq:height}) and (\ref{eq:functional})): 
\begin{equation}
	\label{eq:defhessian}
	{\rm Hess} \, F \, = \, \left(
		\begin{array}{cc}
			\frac{\partial^2 F}{\partial (\partial_x h)^2}  &  \frac{\partial^2 F}{\partial (\partial_x h) \partial (\partial_y h)} \\
			\frac{\partial^2 F}{\partial (\partial_y h) \partial (\partial_x h)}  &   \frac{\partial^2 F}{\partial (\partial_y h)^2}
		\end{array}
	\right) \, = \,
	\frac{1}{\pi^2} \left(
		\begin{array}{cc}
			\frac{\partial^2 F}{\partial m_x^2}  &  -\frac{\partial^2 F}{\partial m_x \partial m_y } \\
			-\frac{\partial^2 F}{\partial m_y \partial m_x}  &   \frac{\partial^2 F}{\partial m_y^2}
		\end{array}
	\right) .
\end{equation}
This leads to the first claim that we intend to check numerically:   \\

\indent	{\it (i) the fluctuations of the height field inside the arctic curve in the interacting six-vertex model are Gaussian, and they are captured by the above action, where ${\rm Hess \,} F$ is the Hessian of the free energy evaluated at $h = h_0$.}  \\

This claim should be very natural for most readers. It is well-known to be true in the free fermion case; here we simply want to check that it carries over to the interacting case.

Since ${\rm Hess \, F}$ is a position-dependent $2\times 2$ (positive) matrix, we may regard $g = ({\rm Hess \, F})^{-1}$ as an emergent metric that equips the domain $\Omega$. Then we define the coupling constant 
\begin{equation}
	K  = \left( 4 \pi \sqrt{{\rm det}({\rm Hess} \, F)}\right)^{-1} , 
\end{equation}
and we rewrite the quadratic action in (\ref{eq:action0}) as
\begin{equation}
\label{eq:action}
	S[\delta h] \, := \, \int \frac{\sqrt{\det g} \, d^2 {\rm x} }{ 8 \pi K }  g^{ab}( \partial_a \delta h )( \partial_b \delta h)\,.
\end{equation}
This leads us to our second claim, which is perhaps less expected than the first one: \\

\indent	{\it (ii) contrary to the free fermion case, where the constant $K$ is quantized ($K$ is always identically one in that case), in the interacting six-vertex model, the coupling constant $K$ becomes position-dependent. This has the direct consequence that the fluctuations are no longer captured by a conformally invariant Gaussian Free Field (GFF) in the domain $\Omega$ equipped with the conformal class of the metric $[g]$. Instead, they are  captured by an GFF that depends both on $[g]$ and on the function ${\rm x} \mapsto K({\rm x})$ on $\Omega$.}  \\

[This {\it inhomogeneous} GFF was recently studied by Y. Brun and one of us in the context of inhomogeneous quantum systems in one spatial dimension \cite{brun2017inhomogeneous}.] To appreciate the difference between the familiar free fermion case, and the interacting case, it is convenient to pick a system of isothermal coordinates $({\rm x}^1, {\rm x}^2)$ for the metric $g$. While in the free fermion case the two-point function (or covariance function) that defines the GFF is simply the Green's function of the Laplacian $\Delta = \nabla^2= \partial^2_{1} + \partial^2_{2}$, in the interacting case it becomes the Green's function of a generalized Laplacian parametrized by the function $K({\rm x})$ \cite{brun2017inhomogeneous},
\begin{equation}
 \nabla_{\rm x} \left( \frac{1}{K({\rm x})} \nabla_{\rm x} \left< h({\rm x}) h({\rm x}') \right> \right) \, = \, -4\pi \delta^{(2)} ({\rm x} - {\rm x}') \,.
\end{equation}

\medskip

The rest of the paper is dedicated to developing a strategy for checking our claims (i) and (ii) above. It is organized as follows. In section \ref{sec:evaluationF}, we present a method that allows us to efficiently calculate the free energy $F(m_x,  m_y)$. The method is based on a numerical solution of the Bethe equations in the presence of a complex twist. In the free-fermion case $\Delta=0$ an analytical expression for $F(m_x,m_y)$ can even be derived, see Eq. \eqref{Fff}. In section \ref{sec:transfermatrix} we provide numerical evidence of our claim (i), namely that the fluctuations of the height field inside the arctic curve are Gaussian, and are given by the action (\ref{eq:action0})--(\ref{eq:action}). Section \ref{sec:K} contains checks of our claim (ii): contrary to the free fermion case ($\Delta =0$), in the interacting case the coupling constant $K$ varies with position. We demonstrate this variation both analytically, by computing $K$ and its first two derivatives at selected points, and by numerically evaluating $K$ throughout the critical region. We find that for $\Delta\leq -1/2$ the function $K$ becomes singular at $m_x,m_y=0$.
We conclude in section \ref{sec:conclusion}. A technical point is relegated to Appendix \ref{app:propermin}.

\section{Evaluation of $F(m_x,m_y)$ from Bethe equations with imaginary extensive twist}
\label{sec:evaluationF}
In this section we derive the free energy $F(m_x,m_y)$ numerically from the Bethe ansatz in the interacting case $\Delta\neq 0$, and analytically when $\Delta=0$.

\subsection{Imaginary twist and the monodromy matrix}

Denote  $Z_{N_x N_y}(m_x,m_y)$ the partition function of the six-vertex model on an $N_x\times N_y$ square lattice with periodic boundary conditions, imposing mean value magnetizations $\langle m_x\rangle=m_x\in [-1,1]$ and $\langle m_y\rangle=m_y \in [-1,1]$. \eqref{eq:height} on the lattice gives the following expression for the mean value of the magnetizations:
\begin{equation}
\begin{aligned}
\langle m_x\rangle&=\frac{\text{card } \{ \text{up arrows} \}-\text{card } \{ \text{down arrows} \}}{  N_x N_y  }\\
\langle m_y\rangle&=\frac{\text{card } \{ \text{right arrows} \}-\text{card } \{ \text{left arrows} \}}{N_x N_y }\,,
\end{aligned}
\end{equation}
so that both belong to $[-1,1]$. The free energy $F(m_x,m_y)$ is defined as 
\begin{equation}
F(m_x,m_y)=\underset{N_x,N_y\to\infty}{\lim} \left( -\frac{1}{N_x N_y}\log Z_{N_x N_y}(m_x,m_y)\right)\,.
\end{equation}
Let us explain how $F(m_x,m_y)$ can be determined with Bethe ansatz techniques. Denote the monodromy matrix $T_{N_x}$ by
\begin{equation}
T_{N_x}=\left( \begin{matrix}
A & B \\
C & D
\end{matrix}\right)\,,
\end{equation}
where $A,B,C,D$ are the $2^{N_x} \times 2^{N_x}$ matrices propagating upwards, with imposed fixed boundary conditions on the left and right boundaries. Specifically, one has right arrows at edges $0$ and $N_x$ for $A$, left arrows at edges $0$ and $N_x$ for $D$, a right arrow at edge $0$ and a left arrow at edge $N_x$ for $B$, a left arrow at edge $0$ and a right arrow at edge $N_x$ for $C$. The usual transfer matrix with periodic boundary conditions is defined as the trace of the monodromy matrix. The transfer matrix conserves the total magnetization $m_x$, so one can work in a fixed magnetization sector, and then take the trace over the $N_y$-th power of the transfer matrix in that sector.

However, the magnetization $m_y$ cannot be imposed that way, since it is not associated to a conserved charge of the row-to-row transfer matrix. We  need to resort to the following trick. Since each $A$ contributes positively to $\langle m_y\rangle$ and $D$ negatively, we introduce an additional parameter $\varphi$ and define the transfer matrix $t_{N_x}$ as
\begin{equation}
\label{transfer matrix}
t_{N_x} =e^{-N_x \varphi}A+e^{N_x \varphi}D\,,
\end{equation}
Denote $Z_{N_x N_y}(m_x,\varphi)$ the partition function of this model with imposed magnetization $\langle m_x\rangle =m_x$. We have
\begin{equation}
\label{partition}
Z_{N_x N_y}(m_x,\varphi)=\sum_{i_1,...,i_{N_y} \in \{-1,1\}^{N_y} }e^{-(i_1+...+i_{N_y}) N_x \varphi } \tr_{m_x} X_{i_{N_y}} \dots X_{i_1}  \,,
\end{equation}
where $X_1=A$, $X_{-1}=D$, and where $\tr_{m_x}$ is the trace on the states with magnetization $m_x$. Denote its free energy $f(m_x,\varphi)$, where 
\begin{equation}
f(m_x,\varphi)=\underset{N_x, N_y\to\infty}{\lim} \left( -\frac{1}{N_x N_y}\log Z_{N_x N_y}(m_x,\varphi)\right)\,.
\end{equation}
In terms of this transfer matrix, the mean value $\langle m_y \rangle$ is given by
\begin{equation}
\langle m_y \rangle = \frac{1}{Z_{N_x N_y}}\sum_{i_1,...,i_{N_y} \in \{-1,1\}^{N_y} }\frac{i_1+...+i_{N_y}}{N_y}e^{-(i_1+...+i_{N_y})N_x \varphi } \tr_{m_x} X_{i_{N_y}} \dots X_{i_1} \,,
\end{equation}
which is exactly
\begin{equation}
\langle m_y \rangle =- \partial_\varphi \frac{1}{N_x N_y}\log Z_{N_x N_y}(m_x,\varphi)\,.
\end{equation}
Thus, in the thermodynamic limit, to impose a magnetization $m_y$ one has to fix $\varphi$ to the value where the corresponding derivative $\partial_\varphi f$ gives $m_y$. In this sense, $\varphi$ and $m_y$ are conjugate variables.
Moreover it implies that, in Eq. \eqref{partition}, the terms with magnetization $m_y=\partial_\varphi f$ dominate in the sum. Then
\begin{equation}
Z_{N_x N_y}(m_x,\varphi)\sim e^{-N_x N_y  m_y \varphi }\, Z_{N_x N_y}(m_x,m_y)\,,
\end{equation}
which gives, in the thermodynamic limit,
\begin{equation}
\label{free energy}
F(m_x,m_y)=f(m_x,\varphi)-\varphi m_y \quad \text{where }\varphi \text{ satisfies} \quad m_y= \partial_\varphi f(m_x,\varphi)\,.
\end{equation}
Thus the free energy $F(m_x,m_y)$ is the Legendre transform of $f(m_x,\varphi)$ in terms of the conjugate variables $m_y$ and $\varphi$.

Next, we observe that $f(m_x,\varphi)$ is the logarithm of the maximal eigenvalue of the transfer matrix $t_{N_x}$ divided by $N_x$ in the sector of magnetization $m_x$. It is well-known when $\varphi=0$ that the eigenvalues of the transfer matrix \eqref{transfer matrix} can be obtained with the Bethe ansatz. The parameter $\varphi$ simply multiplies the monodromy matrix elements $A$ and $D$, so that similar Bethe ansatz equations can be found at $\varphi\neq 0$. Defining $\gamma$ by
\begin{equation}
c =  2 \cos \frac{\gamma}{2},
\end{equation}
 (see Fig. \ref{fig:6v} for the definition of the parameter $c$) or equivalently by
\begin{equation}
\Delta=-\cos\gamma\,,
\end{equation}
the eigenvalues of \eqref{transfer matrix} read (for even sizes $N_x$), up to a global multiplicative factor,
\begin{equation}
\label{su2}
 \Lambda=e^{-N_x \varphi}\prod_{j=1}^{M}\dfrac{\sinh(\lambda_j+i\gamma)}{\sinh(\lambda_j)}+e^{N_x \varphi}\prod_{j=1}^{M}\dfrac{\sinh(\lambda_j-i\gamma)}{\sinh(\lambda_j)}\,,
\end{equation} 
where the $M$ numbers $\lambda_1,...,\lambda_L$ are the roots of the Bethe equations
\begin{equation}
\label{be2}
\left(\dfrac{\sinh(\lambda_i+i\gamma/2)}{\sinh(\lambda_i-i\gamma/2)} \right)^{N_x}=e^{2 N_x \varphi}\prod_{j\neq i}\dfrac{\sinh(\lambda_i-\lambda_j+i\gamma)}{\sinh(\lambda_i-\lambda_j-i\gamma)}\,.
\end{equation}
Notice that in the usual treatment of the twisted XXZ spin chain \cite{AlcarazBarberBatchelor2}, the twist factor that is introduced in the Bethe equations reads instead $e^{i\tilde{\varphi}}$. There are two important differences in our case: the twist is imaginary and it scales extensively with $N_x$: $\tilde{\varphi}=-2iN_x \varphi$. The Bethe ansatz techniques however remain applicable in spite of these changes.

The number of roots $M$ in \eqref{su2}--\eqref{be2} is linked to the magnetization $m_x$ through
\begin{equation}
\label{L}
M=  N_x \frac{1- m_x}{2}\,.
\end{equation}
The Bethe ansatz enables us to compute  the function $F(m_x,m_y)$ with high precision, by numerically solving the Bethe equations \eqref{be2}. \\

To conclude this section, we note that, since we restrict to the isotropic six-vertex model in this paper (see Fig. \ref{fig:6v}), the free energy $F(.,.)$ must be symmetric, $F(m_x, m_y) = F(m_y, m_x)$. However, in the way we calculate $F(.,.)$, the two components of the magnetization are treated very differently, thus it is a highly non-trivial check of the validity of our numerical evaluation of $F(.,.)$ that the result should be symmetric. 

At the particular point that maps to free fermions ($\Delta = 0$), a proof that $F(m_x,m_y)$, as defined in (\ref{free energy}), is indeed symmetric, is given in Section~\ref{sec:symm_free_energy} below.

\subsection{Numerical results for the free energy $F(m_x, m_y)$ at $\Delta \neq 0$}
\label{sec:roots}
It is well-known that, in the absence of an imaginary twist $\varphi$, the Bethe ansatz equations \eqref{be2} lead to real roots $\lambda_i$ and for the sector of magnetization $m_x=0$ the root density can be computed analytically, see e.g. Ref. \cite{AlcarazBarberBatchelor}. However, as soon as the imaginary twist $\varphi$ is non-zero, the Bethe roots acquire an imaginary part and are situated on a curve in the complex plane in the thermodynamic limit. When $\varphi\to \pm \infty$, they gather around $\pm i \gamma/2$. The shape of this curve is not known, nor is the root density. The plot \ref{figroots} shows the Bethe root configuration for the ground state for $\gamma = \pi/5$ (i.e. $c= 2 \cos \frac{\pi}{10}$, or $\Delta = - \cos \frac{\pi}{5}$) at $N_x =156$ for different values of twist.
 \begin{figure}[H]
 \begin{center}
\includegraphics[scale=0.6]{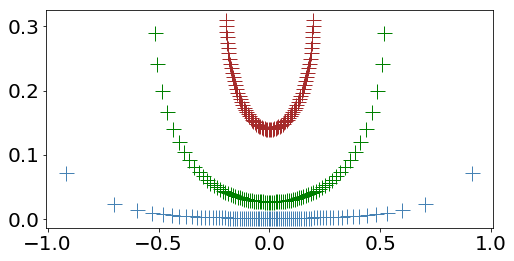} 
\end{center}
\caption{Distribution of Bethe roots in the complex plane for the largest eigenvalue of the twisted transfer matrix $t_{N_x}$ at $\gamma=\pi/5$, i.e.  $\Delta\approx -0.89$, for $N_x =156$ and $\varphi=0.01$ (blue), $\varphi=0.2$ (green), $\varphi=1$ (red).}
 \label{figroots}
\end{figure}
Once the function $f(m_x,\varphi)$ is known, the free energy $F(m_x,m_y)$ can be computed numerically. One finds a similar qualitative behaviour of the function for different values of $\gamma$ ($\Delta$ positive or negative), see Fig. \ref{fig:F}. Although the magnetizations $m_x$ and $m_y$ are treated very differently, the symmetry of the function $F(m_x,m_y)$ is very good: the relation $F(m_x,m_y)=F(m_y,m_x)$ is satisfied up to a relative error of order $10^{-5}$, with the calculation scheme developed above with $N_x=156$. The error comes from the fact that we are not in the thermodynamic limit.
As pointed out in the previous subsection, checking this symmetry provides a rather stringent test of the correctness of the numerical method.

 \begin{figure}[H]
 \begin{center}
\includegraphics[scale=0.6]{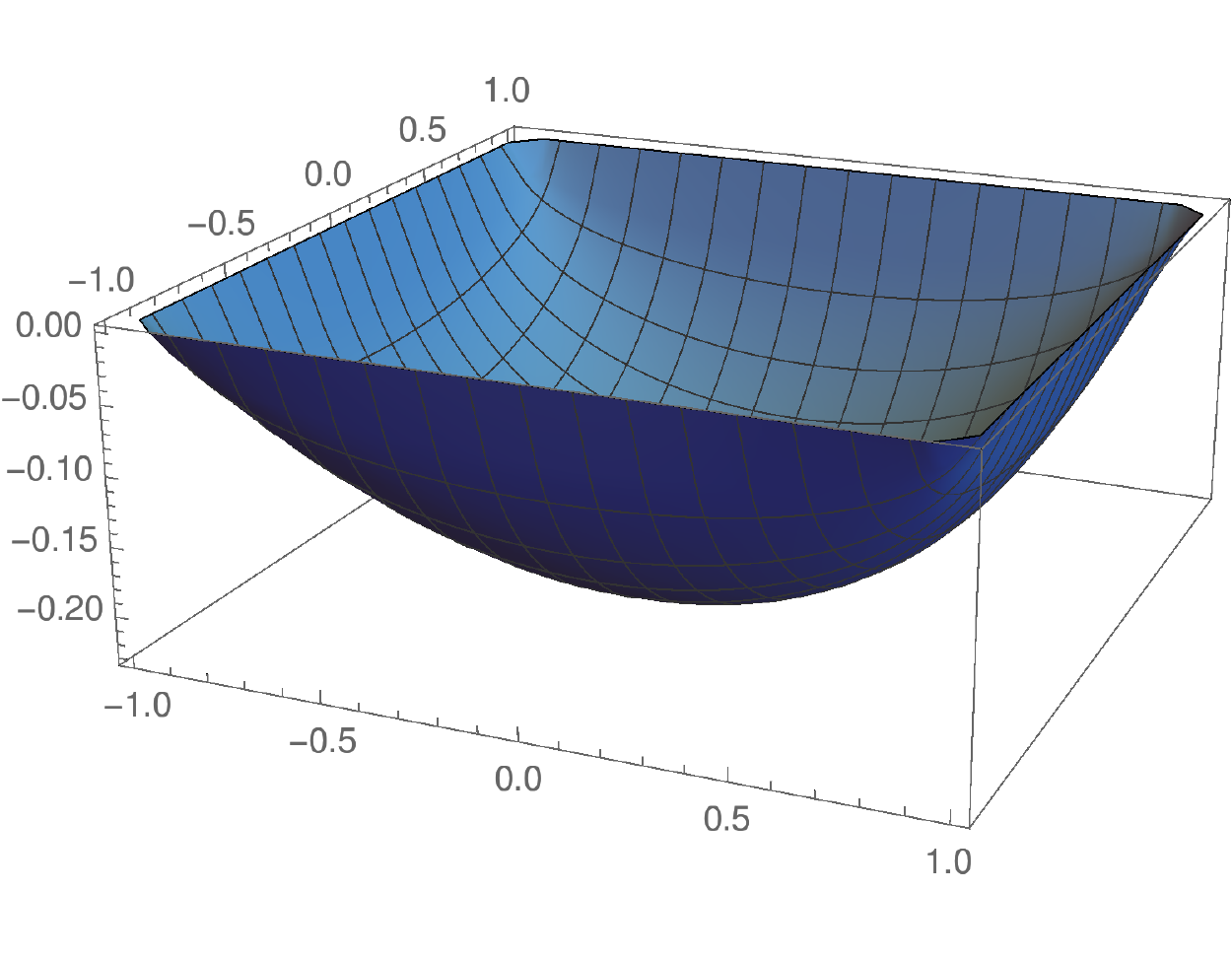} 
\includegraphics[scale=0.6]{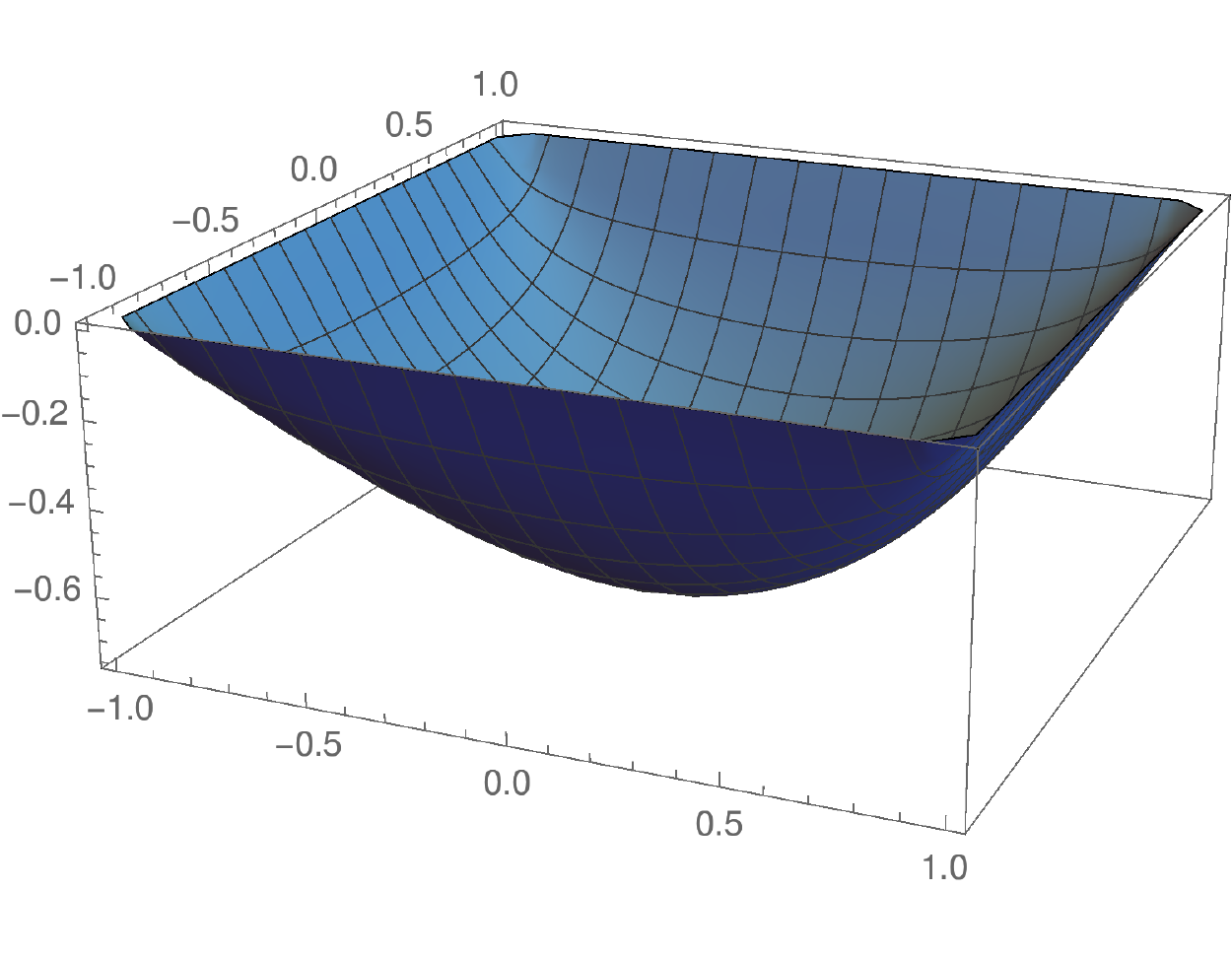} 
\end{center}
\caption{Free energy $F(m_x,m_y)$ as a function of $m_x,m_y$ for $c=1/2$, i.e. $\Delta=0.875$ (left) and $c=19/10$, i.e. $\Delta\approx -0.8$ (right).}
\label{fig:F}
\end{figure}

\subsection{The example of the free fermion point ($\Delta = 0$)}
\label{sec:symm_free_energy}
In this subsection we apply our transfer matrix method for the evaluation of $F(m_x,m_y)$ to the free fermion case, where an analytical calculation is  possible. We express the final result for the free energy $F(m_x,m_y)$, as calculated from Eq. \eqref{free energy}, in a way that is clearly symmetric in $m_x$ and $m_y$.

\subsubsection*{The free energy $f(m_x,\varphi)$}

To compute the free energy $F(m_x,m_y)$ we need the free energy $f(m_x,\varphi)$, imposing $m_x$ but not $m_y$, and with the parameter $\varphi$. The free fermion case corresponds to $\gamma=\pi/2$, or equivalently $c=\sqrt{2}$ or $\Delta = 0$. In this case the Bethe equations are
\begin{equation}
\left(\dfrac{\sinh(\lambda_i+i\pi/4)}{\sinh(\lambda_i-i\pi/4)} \right)^{N_x}=e^{2{N_x}\varphi}\,,
\end{equation}
whose solutions are
\begin{equation}
\lambda_k=\argth \tan (k\pi/{N_x}+i\varphi)\quad\,,
\end{equation}
where $k$ is an integer (if $N_x$ is even) or a half-integer (if $N_x$ is odd) between $-N_x/2$ and $N_x/2$. Imposing a magnetization $m_x$ corresponds to considering only $M$ of these roots, with $M$ given by \eqref{L}. The maximal eigenvalue with this constraint is obtained by taking $k$ between $-M/2$ and $M/2$. The log of the modulus of this eigenvalue, determined by \eqref{su2}, becomes in the thermodynamic limit
\begin{equation}
f(m_x,\varphi) \, =  \, -|\varphi|-\Re \int_{-(1- m_x)/4}^{(1-m_x)/4} \log \frac{\sinh( \argth \tan (\pi x +i\varphi)+i\pi/2)}{\sinh \argth \tan (\pi x +i\varphi)}  dx\,,
\end{equation}
which simplifies to
\begin{equation}
\label{f}
f(m_x,\varphi) \, = \, - |\varphi| + \Re \int_{-(1-m_x)/4}^{(1- m_x)/4} \log \tan (\pi x +i\varphi) dx\,.
\end{equation}
Differentiating with respect to $\varphi$, we get
\begin{equation}
\begin{aligned}
\partial_\varphi f&=  - \text{sign} \varphi + \Re \int_{-(1-m_x)/4}^{(1- m_x)/4} \partial_\varphi\log \tan (\pi x +i\varphi) dx\\
&=  - \text{sign} \varphi +  \Re \frac{i}{\pi}\int_{-(1-m_x)/4}^{(1- m_x)/4} \partial_x\log \tan (\pi x +i\varphi) dx\\
&=  - \text{sign} \varphi - \frac{1}{\pi}\Im \log \frac{\tan ( \tfrac{\pi}{4}(1-m_x) +i\varphi)}{\tan ( -\tfrac{\pi}{4}(1-m_x) +i\varphi)}\,.
\end{aligned}
\end{equation}
Using the fact that the imaginary part of $\log (a+ib)$ is $\arctan b/a$ when $a>0$, and using standard trigonometric identities, one gets after some algebra
\begin{equation}
\partial_\varphi f= - \frac{2}{\pi}\arctan \dfrac{\sinh 2\varphi}{\cos  \pi m_x/2}\,.
\end{equation}
Using now
\begin{equation}
\label{fmy}
\partial_\varphi f= m_y\,,
\end{equation}
we get the following relation between the parameter $\varphi$ and the magnetizations
\begin{equation}
\label{phi}
\sinh 2\varphi= -\cos \frac{\pi m_x}{2}\tan \frac{ \pi m_y}{2}\,.
\end{equation}
The free energy $F(m_x,m_y)$ is then, in terms of $f(m_x, \varphi)$,
\begin{equation}
F(m_x,m_y)=f(m_x,\varphi(m_x,m_y)) - \varphi(m_x,m_y)m_y\,,
\end{equation}
where the function $\varphi(m_x,m_y)$ is given by \eqref{phi}.

\subsubsection*{Symmetry of the free energy $F(m_x,m_y)$ in the isotropic six-vertex model}

Let us now rewrite $F(m_x,m_y)$ in a form that is manifestly symmetric under exchange of $m_x$ and $m_y$. First we observe that because of the relation \eqref{fmy} we have 
\begin{equation}
\partial_{m_y} F(m_x,m_y)=- \varphi(m_x,m_y)\,,
\end{equation}
so that in the following variables:
\begin{equation}
	\label{eq:ab}
\alpha=\sin \frac{\pi m_x}{2},\quad \beta=\sin \frac{\pi m_y}{2}\,,
\end{equation}
the cross derivative of $F$ reads
\begin{equation}
\partial_{m_x}\partial_{m_y}F(m_x,m_y)= -\frac{\pi}{4}\dfrac{\alpha\beta}{\sqrt{1-\alpha^2\beta^2}}\,,
\label{F_cross_derivative} 
\end{equation}
which is clearly symmetric in $m_x,m_y$. A first integration gives
\begin{equation}
\partial_{m_y}F(m_x,m_y)= -\frac{\pi}{4}\int_0^{m_x} \dfrac{\alpha(u)\beta(m_y)}{\sqrt{1-\alpha^2(u)\beta^2(m_y)}}du  +  \frac{1}{2}\argsh \tan \frac{\pi m_y}{2}\,,
\end{equation}
and a second one
\begin{equation}
F(m_x,m_y)= -\frac{\pi}{4}\int_0^{m_y} \int_0^{m_x}  \dfrac{\alpha(u)\beta(v)}{\sqrt{1-\alpha^2(u)\beta^2(v)}}du dv  + \frac{1}{2}\int_0^{m_y}\argsh \tan \pi \frac{ v}{2}dv + F(m_x,0)\,.
\end{equation}
Note now that $F(m_x,0)=f(m_x,0)$ given in \eqref{f}. After a change of variable $x=(1-u)/4$ and some manipulations, one gets
\begin{equation}
F(m_x,0)=F(0,0) +  \int_0^{m_x}\argth \tan \pi \dfrac{u}{4}du\,.
\end{equation}
We now make use of the following relations, readily obtained by computing their derivatives
\begin{equation}
\begin{aligned}
\argth \tan \tfrac{x}{2}&=\tfrac{1}{2}\argsh \tan x\\
\argsh \tan &= \argth \sin\,,
\end{aligned}
\end{equation}
to get
\begin{equation}
\label{Fff}
\begin{aligned}
F(m_x,m_y)&= - \dfrac{1}{\pi}\int_0^{\alpha}\int_0^{\beta}\dfrac{xy}{\sqrt{(1-x^2y^2)(1-x^2)(1-y^2)}}dxdy   \\
&+ \frac{1}{\pi}\left(\int_0^\alpha \frac{\argth x}{\sqrt{1-x^2}} dx+\int_0^\beta \frac{\argth y}{\sqrt{1-y^2}}dy \right)+F(0,0) %\\
%&\text{where}\quad a=\sin \pi m_x/2,\quad b=\sin  \pi m_y/2\,,
\end{aligned}
\end{equation}
where $\alpha(m_x)$ and $\beta(m_y)$ are given by Eq. (\ref{eq:ab}), which shows that $F(m_x,m_y)$ is indeed symmetric.

\newpage

\section{Gaussianity of fluctuations inside the arctic curve}
\label{sec:transfermatrix}

Armed with the numerically efficient procedure for calculating the free energy $F(m_x, m_y)$ presented in section \ref{sec:evaluationF}, we now report a number of numerical checks of our first claim, namely that the fluctuations inside the arctic curve are Gaussian, and are captured by the action (\ref{eq:action0}) or equivalently (\ref{eq:action}). The main tool in this section is a numerically exact transfer matrix evaluation of the connected correlations $\left<  h({\rm x}) h({\rm x}') \right>$ on a $20 \times 20$ lattice.

\subsection{Check of Wick's theorem on a $20 \times 20$ lattice}

We study the correlations of the height field $h$ on a finite $N\times N$ lattice with domain-wall boundary conditions. We define the mean value
of $h$ on the lattice,
\begin{equation}
	h_N ({\rm x}) = \langle h({\rm x}) \rangle ,
\end{equation}
see also Figs. \ref{figcolomo} and \ref{figheight}. Our goal is to provide evidence that the action (\ref{eq:action0}) is correct, and we start by checking that the fluctuations of the height field $h({\rm x})$ around its mean value $h_N({\rm x})$ become Gaussian in the thermodynamic limit. To do so, we check that the four-point function of $\delta h({\rm x}) = h({\rm x}) - h_N({\rm x})$ satisfies Wick's theorem,
\begin{eqnarray}
%\begin{aligned}
\langle  \delta h({\rm x}_1) \delta h({\rm x}_2) \delta h({\rm x}_3) \delta h({\rm x}_4) \rangle &=&
\langle    \delta h({\rm x}_1) \delta h({\rm x}_2)  \rangle \langle   \delta h({\rm x}_3) \delta h({\rm x}_4)  \rangle  \\
&+& \langle   \delta h({\rm x}_1) \delta h({\rm x}_3)  \rangle \langle   \delta h({\rm x}_2) \delta h({\rm x}_4)  \rangle +\langle  \delta h({\rm x}_1) \delta h({\rm x}_4)  \rangle \langle   \delta h({\rm x}_2) \delta h({\rm x}_3)  \rangle . \nonumber
%\end{aligned}
\end{eqnarray}
Numerically, we evaluate the following ratio,
\begin{equation}
	\delta W  =  \frac{\left< \phi_1 \phi_2 \phi_3 \phi_4 \right> - \left< \phi_1 \phi_2 \right>\left< \phi_3 \phi_4 \right> - \left< \phi_1 \phi_3 \right>\left< \phi_2 \phi_4 \right> -\left< \phi_1 \phi_4 \right>\left< \phi_2 \phi_3 \right> }{\left< \phi_1 \phi_2 \phi_3 \phi_4 \right>}
\end{equation}
%\begingroup\makeatletter\def\f@size{6}\check@mathfonts
%\begin{equation}
%\delta W=\frac{\langle  \delta h_1 \delta h_1    \rangle-\langle \phi(x_1)\phi(x_2)\rangle \langle \phi(x_3)\phi(x_4)\rangle-\langle \phi(x_1)\phi(x_3)\rangle \langle \phi(x_2)\phi(x_4)\rangle -\langle \phi(x_1)\phi(x_4)\rangle \langle \phi(x_2)\phi(x_3)\rangle}{\langle \phi(x_1)\phi(x_2)\phi(x_3)\phi(x_4)\rangle}
%\end{equation}
%\endgroup
with $\phi_j = \delta h({\rm x}_j)$, which should be zero if Wick's theorem is satisfied. We work on an $N\times N$ lattice, with the maximal value $N=20$. The correlation functions are calculated with a transfer matrix method which is numerically exact, see below for a brief description of the method. There are $N^8$ four-point functions in total, and obtaining all of them is beyond reach for a system size $N=20$, so we restrict ourselves to points $x_i$ located in one of the four central squares of size $N/4$ (i.e., those defined by $N/2\leq x \leq 3N/4$, $N/2\leq y \leq 3N/4$ for the first one, $N/2\leq x \leq 3N/4$, $N/4\leq y \leq N/2$ for the second one, etc). Fig.  \ref{fig:wick} shows normalized histograms of the numerical calculation of these $\delta W$'s.

 \begin{figure}[H]
 \begin{center}
\includegraphics[scale=0.4]{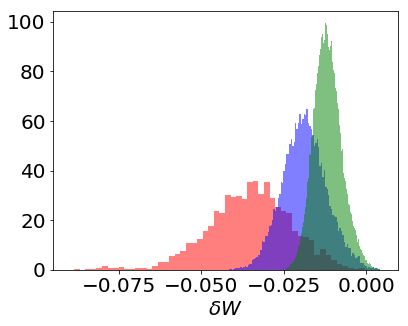}
\includegraphics[scale=0.4]{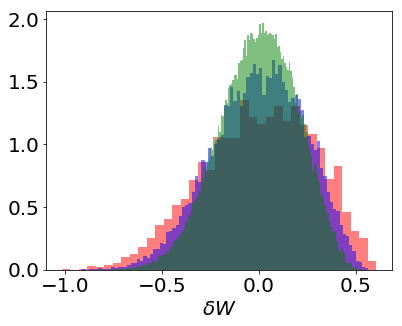}
\includegraphics[scale=0.4]{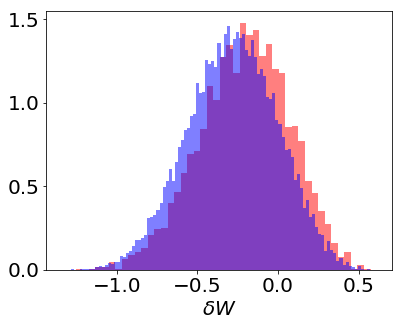}
\includegraphics[scale=0.4]{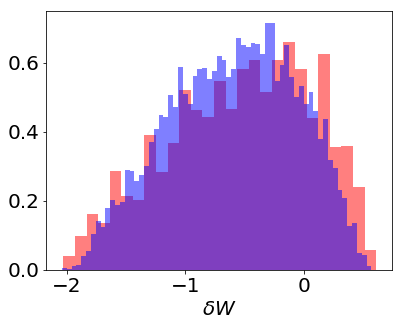}
\end{center}
\caption{Top: histograms of $\delta W$'s for $\phi=\delta h$ for $c=1/2$, i.e. $\Delta=0.875$ (left) and $c=19/10$, i.e. $\Delta\approx -0.8$ (right) in sizes $N=12$ (red), $N=16$ (blue), $N=20$ (green). Bottom: histograms of $\delta W$'s for $\phi=(\delta h)^2$ for $c=1/2$, i.e. $\Delta=0.875$ (left) and $c=19/10$, i.e. $\Delta\approx -0.8$ (right) in sizes $N=12$ (red), $N=16$ (blue).}
\label{fig:wick}
\end{figure}

In Fig. \ref{fig:wick}, the height of a narrow rectangle above the abscissa $x$ is proportional to the number of measured four-point functions whose $\delta W$ is between $x$ and $x+dx$. They are normalized so that the integral is one.  Thus, if Wick's theorem is perfectly satisfied, this plot should be a Dirac delta at $0$. We see on the left that as $N$ increases, the curve becomes narrower and closer to $0$, indicating a convergence toward this distribution as $N\to\infty$. The convergence is much faster for $\Delta>0$ than for $\Delta<0$. As a control, to have an idea of the order of magnitude of $\delta W$ for a generic quantity $\phi$, we also plotted the same histograms for $\phi_j = (\delta h ({\rm x}_j))^2$. The $\delta W$'s are much larger, as it should be since the square of a gaussian random variable is not gaussian and should not satisfy Wick's theorem.

From these numerical results we conclude that the theory describing the fluctuations inside the arctic curve is Gaussian.

\subsubsection*{Brief description of the transfer matrix evaluation of correlation functions}

% proceed as follows. For notational convenience, we call $H_N=-\nabla \cdot (\text{Hess }F \circ h_0) \nabla$ the discrete differential operator on the $N\times N$ lattice, which is also an $N^2\times N^2$ matrix. According to Eq. (\ref{eq:check1}), what we need to check is that this matrix is the inverse of $G_N$, up to subleading finite-size corrections,
%\begin{equation}
%H_N G_N  \, \underset{N\rightarrow \infty}{=} \, \text{Id}_{N^2} \, + \, o (N) .
%\end{equation}
%There are two steps needed to recover the differential operator $H$: first numerically compute the matrix $G_N$, and then analyse the matrix $H_N$ to deduce the dominant form of the differential operator $H$.

We briefly explain our transfer matrix calculation. At each step, the transfer matrix builds one of the $N^2$ vertices, say from bottom to top and left to right, and records the weight of the configurations built up to that point, by labelling them only with the $N$ last vertices that were built. The boundary conditions are implemented at the first and last step, and at the beginning and end of each row. This process yields the partition function, and to have the mean values $\langle h(x_0,y_0)\rangle$ or the correlations $\langle h(x,y) h(x_0,y_0)\rangle$ one simply has to multiply the weights by the corresponding heights when the vertices $(x,y)$ and $(x_0,y_0)$ are built. This method allows for an exact evaluation of the $N^4$ two-point functions on a lattice of size up to $20\times20$, without any statistical error. A Monte Carlo approach would permit larger sizes, but with statistical errors to deal with. In Fig. \ref{figheight} is plotted the mean value of the height function $\langle h_{x,y}\rangle$ and the correlations with point $(-5,-5)$ as an example. In Fig. \ref{figcolomo} is plotted a contour plot of the height function, together with its Colomo-Pronko arctic curve.
 
 \begin{figure}[H]
 \begin{center}
\includegraphics[scale=0.45]{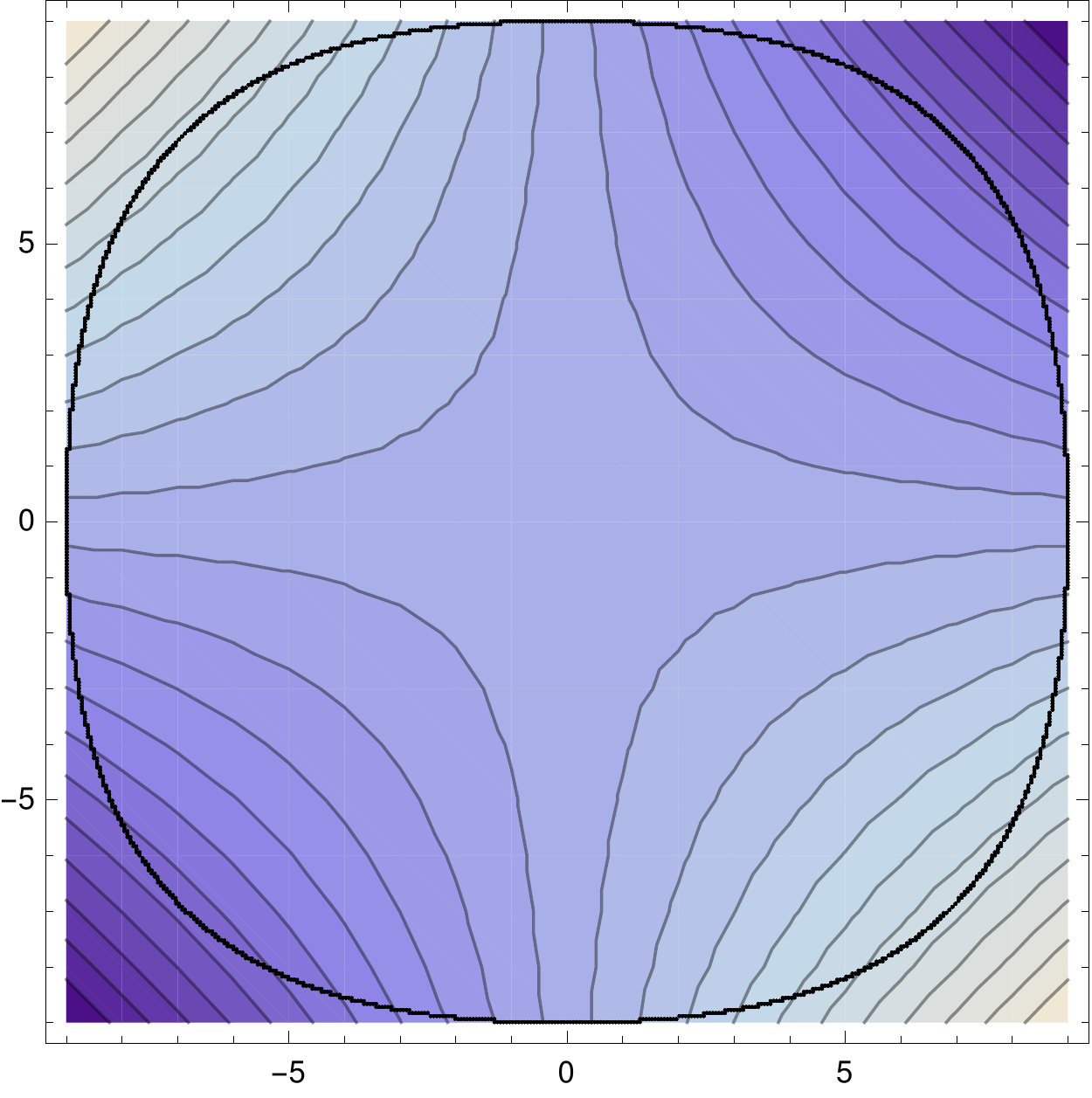} 
\includegraphics[scale=0.45]{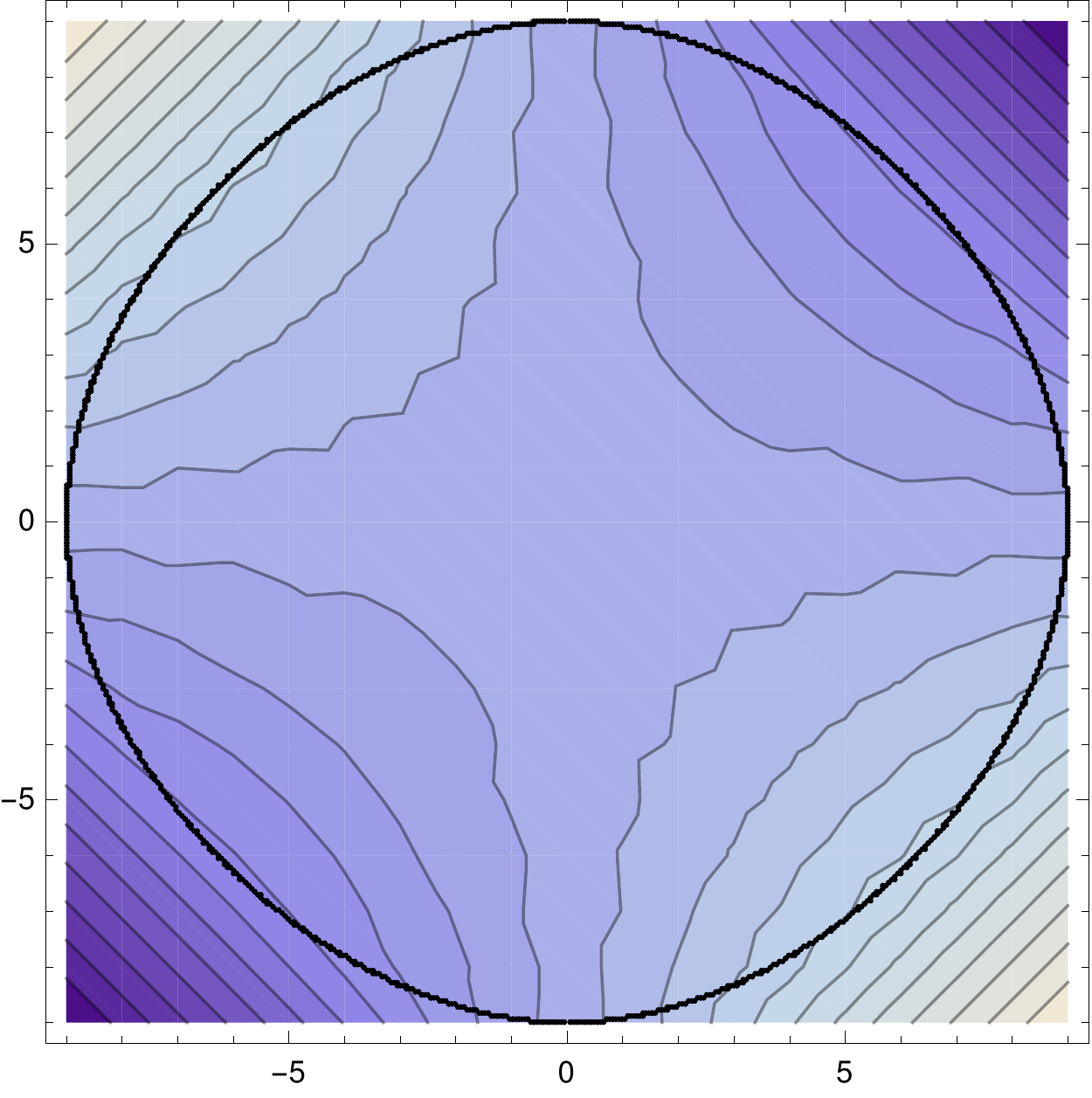} 
\end{center}
\caption{Contour plots of the height function on the $20\times 20$ lattice obtained from a numerically exact transfer matrix calculation. We superimpose the Colomo-Pronko arctic curve \cite{ColomoPronko} (in black). The results are shown for $c=1/2$, i.e. $\Delta=0.875$ (left) and $c=19/10$, i.e. $\Delta\approx-0.8$ (right).}
 \label{figcolomo}
\end{figure}

\subsection{Check of the action (\ref{eq:action0}) on a $20 \times 20$ lattice}

Next, we want to check that the two-point function of the height field is the one predicted by the action \eqref{eq:action0} in the thermodynamic limit. On the lattice, we define the connected correlation matrix, which is an $N^2 \times N^2$ matrix, as
\begin{equation}
G_N({\rm x},{\rm x}')=\langle h_N({\rm x}) h_N({\rm x}')\rangle-\langle h_N({\rm x}) \rangle \langle h_N({\rm x}')\rangle\, ,
\end{equation}
and its corresponding continuous function in the thermodynamic limit
\begin{equation}
G({\rm x},{\rm x}')=\langle h({\rm x}) h({\rm x}')\rangle-\langle h({\rm x}) \rangle \langle h({\rm x}')\rangle\, ,
\end{equation}
see Fig. \ref{figheight}. Then we make the following observation. If the fluctuations of $h$ are described by the action \eqref{eq:action0}, then $h({\rm x})$ and $G  ({\rm x},{\rm x}')$ should satisfy the differential equation 
\begin{equation}
	\label{eq:check1}
-\nabla (\text{Hess }F (h({\rm x}))\nabla )\cdot G({\rm x}, {\rm x}')\,= \,  \delta(x-x'),
\end{equation}
Denote $\mathcal{H}$ the differential operator $\mathcal{H}(h)=-\nabla (\text{Hess }F (h({\rm x}))\nabla )$. In finite size, the matrix $G_N$ should satisfy a similar but discretized version of this differential equation. Denoting $\mathcal{H}_N$ the discretized version of $\mathcal{H}$ in size $N$ (it acts on vectors with $N^2$ variables, so it is also an $N^2\times N^2$ matrix), we have
\begin{equation}
	\label{eq:check1bis}
\mathcal{H}_N(h_N) G_N=I_{N^2}
\end{equation}
As $N\to\infty$, $\mathcal{H}_N$ should be close to $-\nabla_N (\text{Hess }F (h_N({\rm x}))\nabla_N )$ where $\nabla_N$ stands now for discrete derivatives with respect to the two components of ${\rm x} = (x,y)$. This is what we intend to check numerically. Notice that Eq. (\ref{eq:check1}) and Eq. (\ref{eq:check1bis}) relate the two finite-size observables $h_N(.)$ and $G_{N}(.,.)$ to the thermodynamic free energy $F(.,.)$ studied in the previous section, so it really is a non-trivial check of the validity of the action (\ref{eq:action0}).
\begin{figure}[H]
 \begin{center}
\includegraphics[scale=0.5]{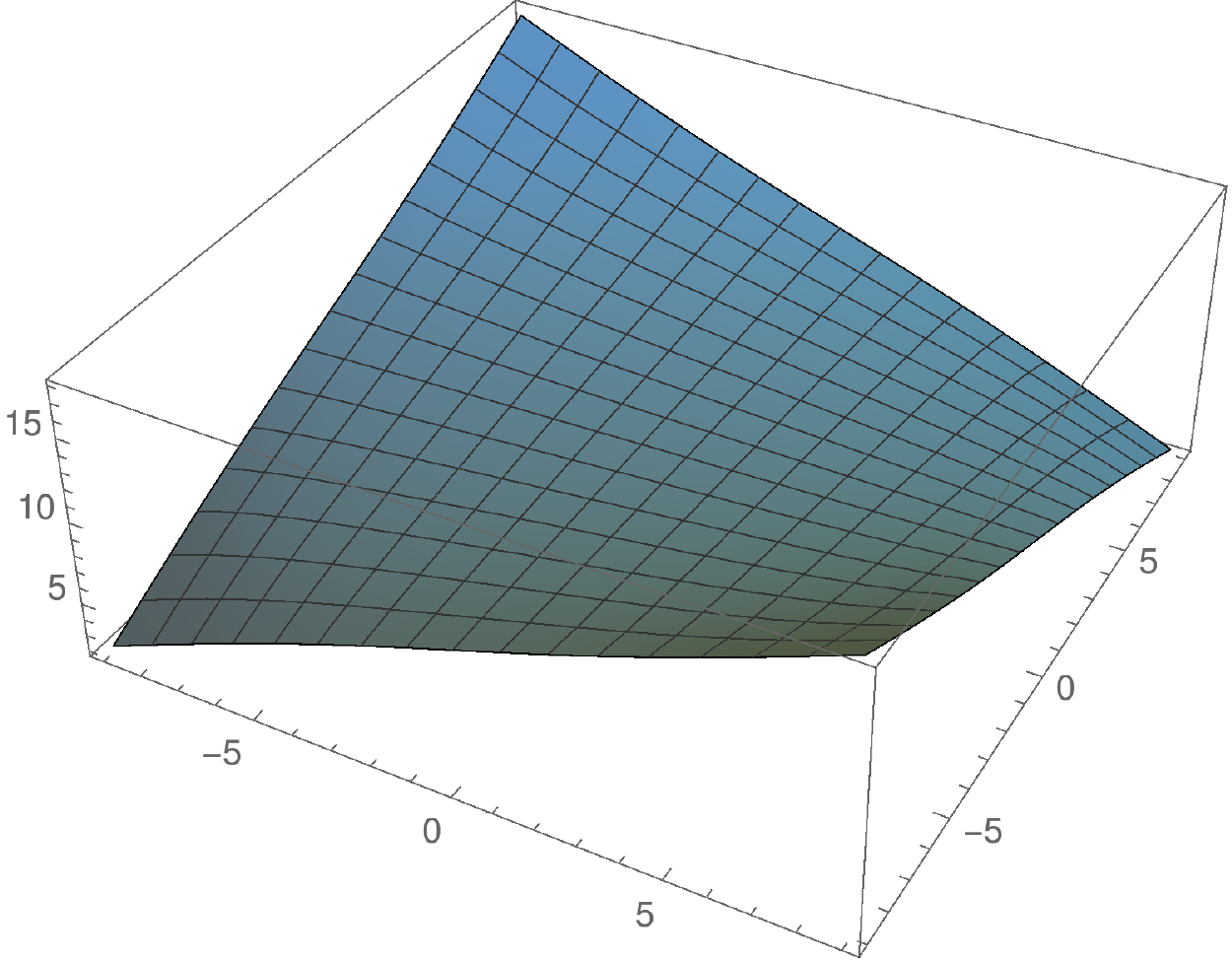}
\includegraphics[scale=0.5]{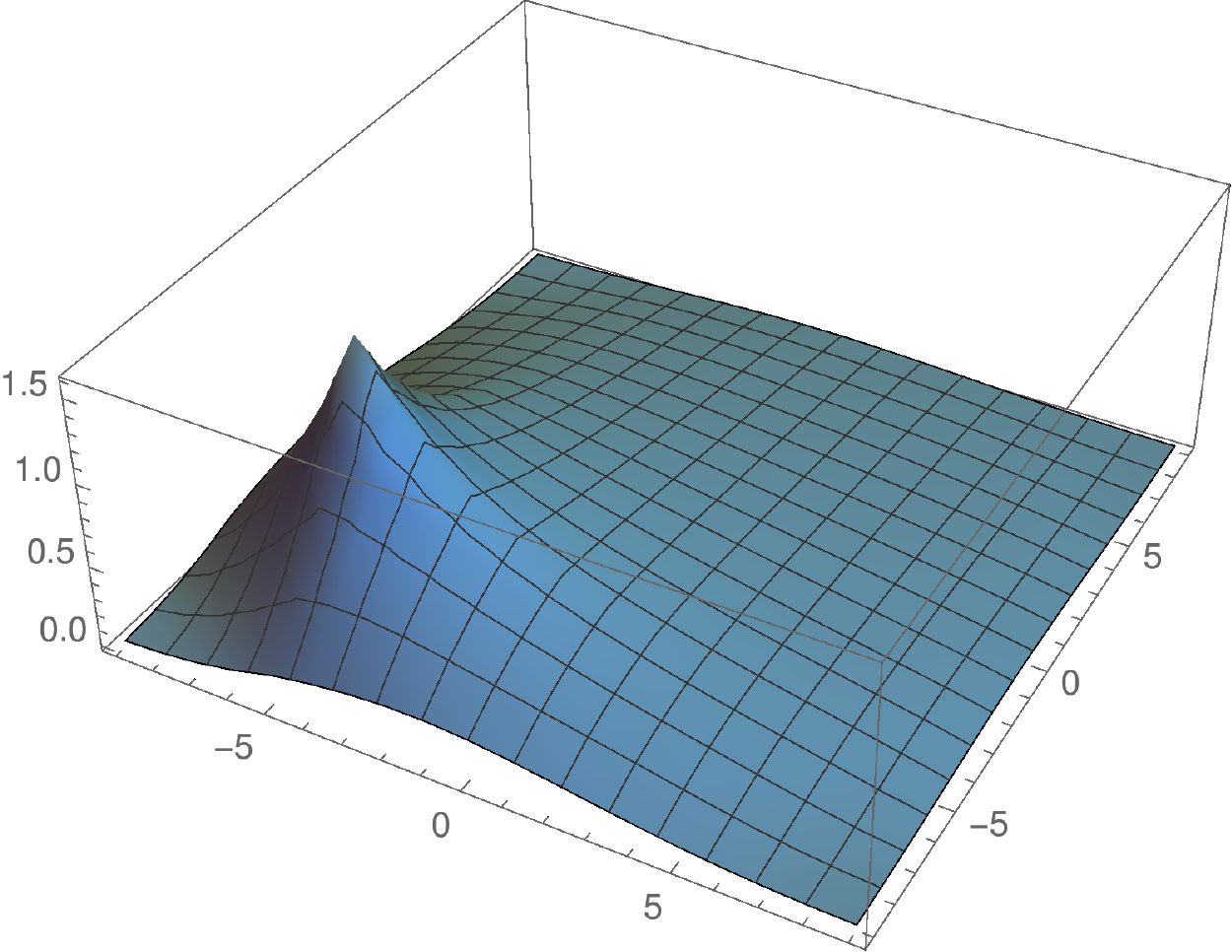}
\end{center}
\caption{Exact lattice one-point and two-point functions of the height field, obtained from a direct transfer matrix evaluation. Left: mean value $h_N(x,y)$ of the height function on the $20\times 20$ lattice. Right: connected correlation $G_N ( (x,y) , {\rm x}_0)$ as a function of $x,y$ on the $20\times 20$ lattice, with ${\rm x}_0 = (x_0,y_0)=(-5,-5)$. Both plots are for $c=1/2$, i.e. $\Delta=0.875$.}
 \label{figheight}
\end{figure}

To show that relation (\ref{eq:check1}) holds, we rely on the following strategy. First, we evaluate the matrix $G_N$ by a numerically exact transfer matrix method, and we invert it. Second, we analyze the matrix $G_N^{-1}$, and provide evidence that, at large $N$, it can be decomposed as a sum of discrete differential operators, the leading one being precisely a discretized version of $-\nabla (\text{Hess }F (h({\rm x}))\nabla )$, as expected from Eq. (\ref{eq:check1}).

\subsubsection*{Expansion of $\mathcal{H}_N$ as a sum of discrete differential operators}
Once the matrix $G_N$ and thus $G_N^{-1}=\mathcal{H}_N$ is known numerically, our goal is to expand it as a sum of discrete differential operators, starting from the lowest-order ones,
\begin{equation}
\mathcal{H}_N= H_1 I_{N^2}  + H_x\partial_x+H_y\partial_y+H_{xx}\partial_x^2+2H_{xy}\partial_x\partial_y+H_{yy}\partial_y^2\, + \, \dots
\end{equation}
Here $I_{N^2}$ is the identity, and $\partial_x$, $\partial_y$, $\partial_x^2 $, $\partial_x\partial_y$, $\partial_y^2$ are discrete differential operators that are all represented as $N^2 \times N^2$ matrices. The corresponding matrices $H_1$, $H_x$, $H_y$, $H_{xx}$, $H_{xy}$, $H_{yy}$ are all {\it diagonal} $N^2 \times N^2$ matrices, with $N^2$ diagonal coefficients associated with the lattice sites.

In principle, there exist different ways to discretize the operators $\partial_x$ and $\partial_y$ by using different finite-difference formulae, so the matrices $H_1,H_x,...$ cannot be read off directly from the coefficients of the matrix $\mathcal{H}_N$. To circumvent that problem, we define $1$ as the vector of size $N^2$ whose components are all $1$, $x$ as the vector of size $N^2$ with components $x_{i,j}=i$ for each $i,j=1,\ldots,N$, and similarly $y_{i,j}=j$, $(x^2)_{i,j}=i^2$, $(y^2)_{i,j}=j^2$ and $(xy)_{i,j}=ij$. In the limit $N\to\infty$ these vectors become the corresponding functions, and regardless of the precise choice of the  discretizations for the operators $\partial_x$ and $\partial_y$, we should have $\partial_x\cdot 1\to 0$ as $N\to\infty$, $\partial_{x}\cdot  x\to 1$, etc. Thus the coefficients of $H$ (now seen as vectors of size $N^2$) can estimated in finite-size by applying the matrix $G_N^{-1}$ on these vectors $1,x,\ldots$:
\begin{equation}
\begin{aligned}
H_1&=G_N^{-1}\cdot 1\\
H_x&=(G_N^{-1}-H_1 \delta_{1,N})\cdot x\\
H_y&=(G_N^{-1}-H_1 \delta_{1,N})\cdot y\\
H_{xx}&=\frac{1}{2}(G_N^{-1}-H_1 \delta_{1,N}-H_x \delta_{x,N})\cdot x^2\\
H_{yy}&=\frac{1}{2}(G_N^{-1}-H_1 \delta_{1,N}-H_y \delta_{y,N})\cdot y^2\\
H_{xy}&=\frac{1}{2}(G_N^{-1}-H_1 \delta_{1,N}-H_x \delta_{x,N}-H_y \delta_{y,N})\cdot xy\,,\\
\end{aligned}
\end{equation}
where $\delta_{1,N}$, $\delta_{x,N}$ and $\delta_{y,N}$ are $N^2\times N^2$ matrices that are discretized versions of the operators $1$, $\partial_x$ and $\partial_y$ without error term of order $2$. We used $\delta_{x,N} f(x)\approx (f(x+\epsilon)-f(x-\epsilon))/(2\epsilon)$.

\subsubsection*{Results}

In order to be compatible with the action (\ref{eq:action0}), this procedure must lead to $H_1=0$, $H_x = \partial_x F\circ h_N$, $H_y=\partial_y F \circ h_N$, $H_{xx}=\partial_x^2 F \circ h_N$, $H_{yy}=\partial_y^2 F \circ h_N$ and $H_{xy}=\partial_x\partial_y F \circ h_N$ in the limit $N \rightarrow \infty$. In Fig. \ref{fig:rawH} we present a direct comparison between the coefficients of the diagonal matrices $H_{xx}$ and $H_{xy}$ obtained from $G_N^{-1}$ with the above procedure, and a numerical evaluation of the discrete derivatives $\partial_x^2 F \circ h_N$ and $\partial_x\partial_y F \circ h_N$. The results are displayed for $c=1/2$ and a lattice of size $20\times 20$ in Figs. \ref{fig:rawH}, \ref{fig:moyH} and \ref{fig:conv}.
 \begin{figure}[H]
 \begin{center}
\includegraphics[scale=0.55]{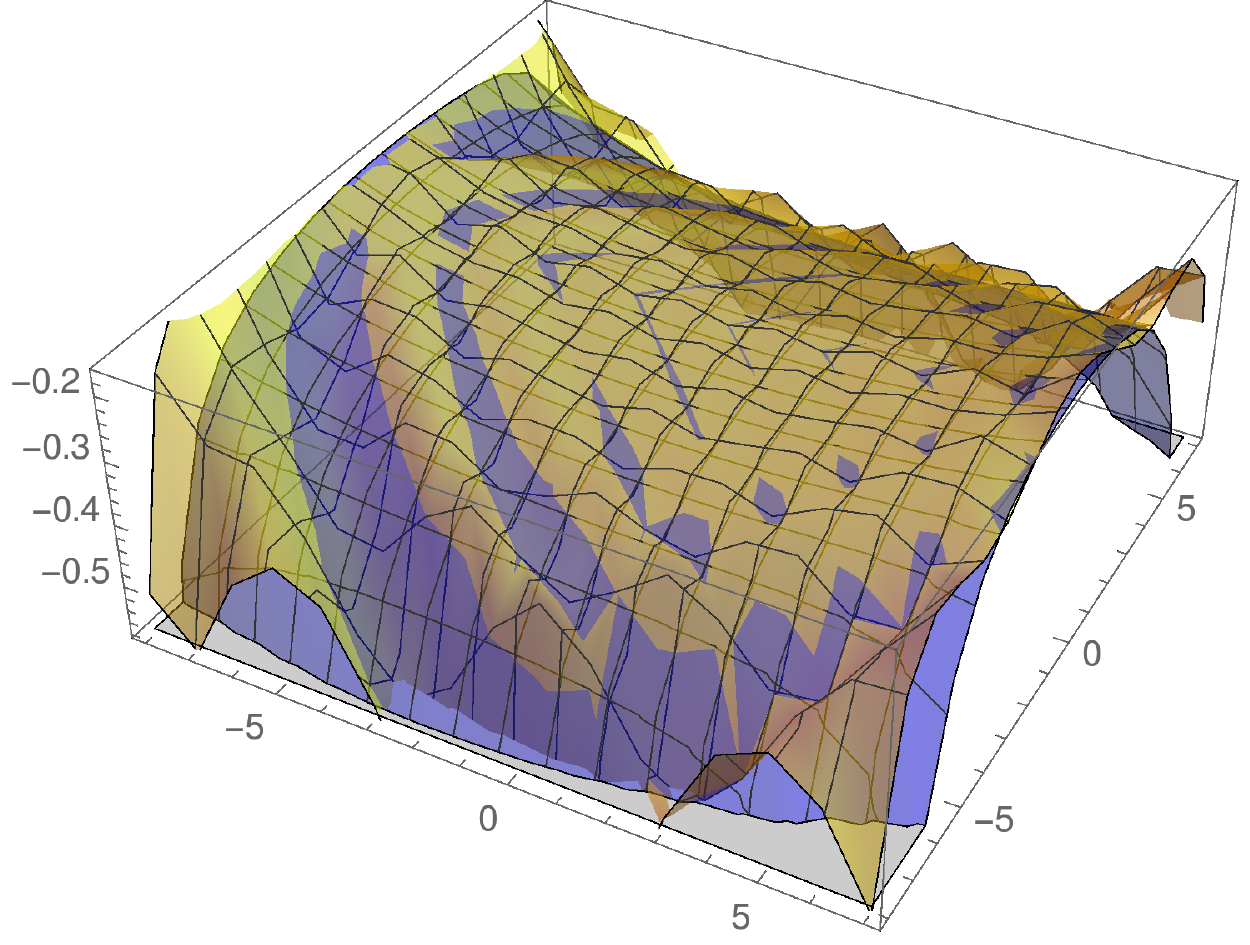} 
\includegraphics[scale=0.55]{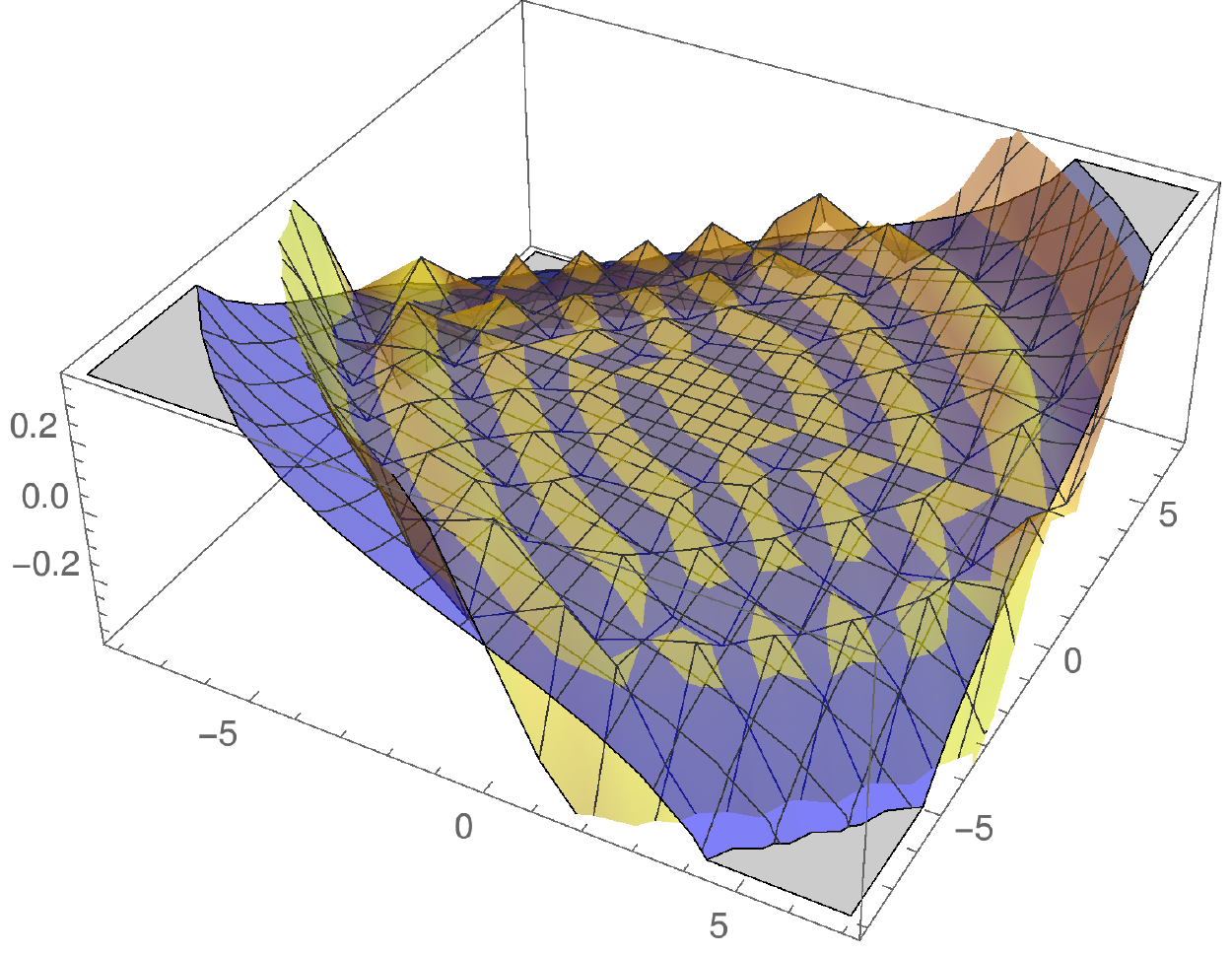} 
\end{center}
\caption{Coefficients of the differential operator $H$ as a function of the position in the $20\times 20$ lattice, with $(0,0)$ being the center (left: $H_{xx}$ ; right: $2H_{xy}$). The transfer matrix measures (in yellow) are compared to the discrete derivatives $\partial_x^2 F \circ h_N$ and $2\partial_x \partial_y F \circ h_N$ with the thermodynamic free energy $F$ calculated from the Bethe ansatz (in blue).}
\label{fig:rawH}
\end{figure}
Figs. \ref{fig:rawH} and \ref{fig:moyH} show the general shape of the coefficients $H_{xx}$ and $H_{xy}$, and illustrate that those computed with the Bethe ansatz are in good agreement with those extracted with the transfer matrix. There are strong parity effects that make the measured curve (in yellow in Fig. \ref{fig:rawH}) oscillate around the theoretical curve (in blue). This is why, in Fig. \ref{fig:moyH}, we show the same data, but where
 \begin{figure}[H]
 \begin{center}
\includegraphics[scale=0.55]{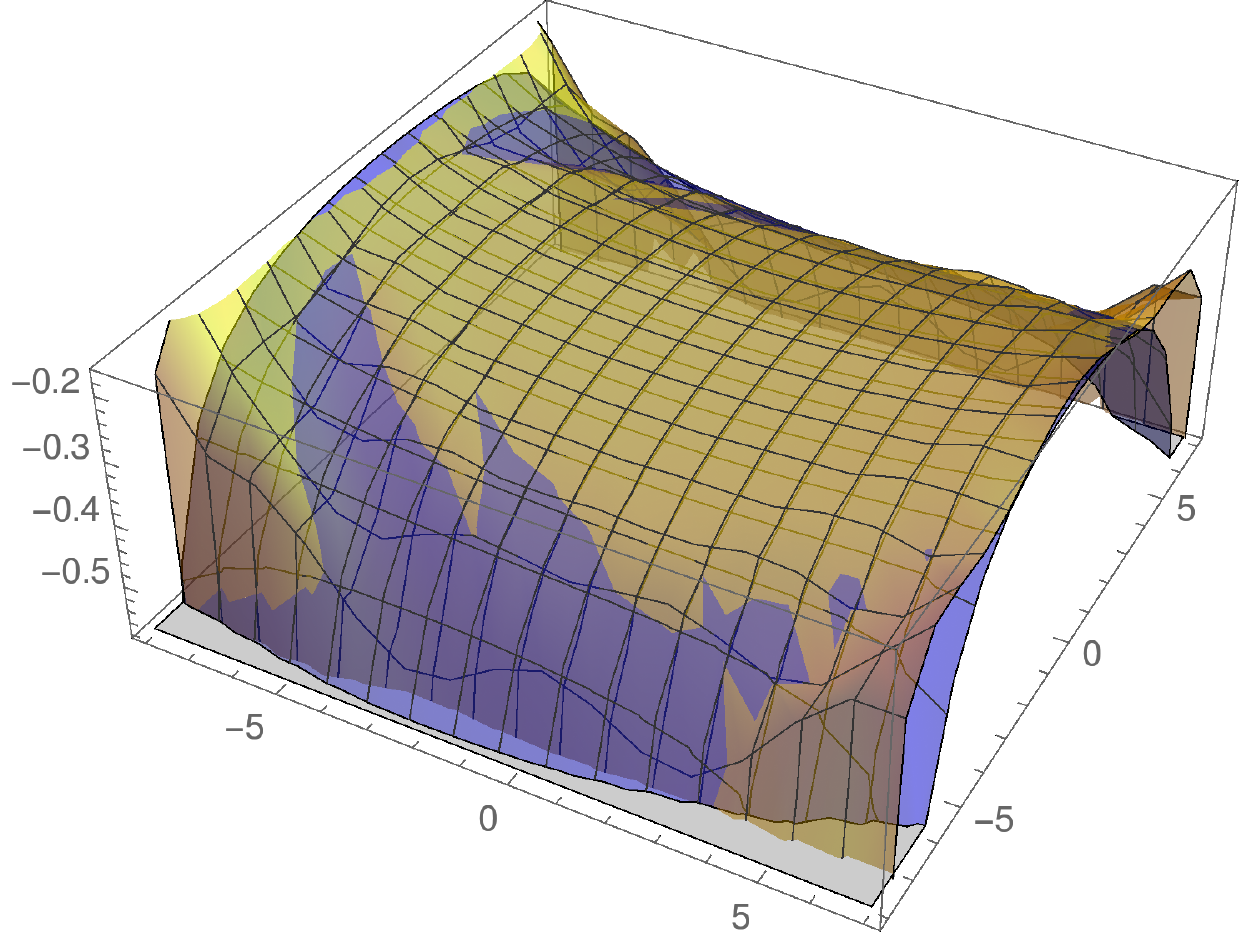}
\includegraphics[scale=0.55]{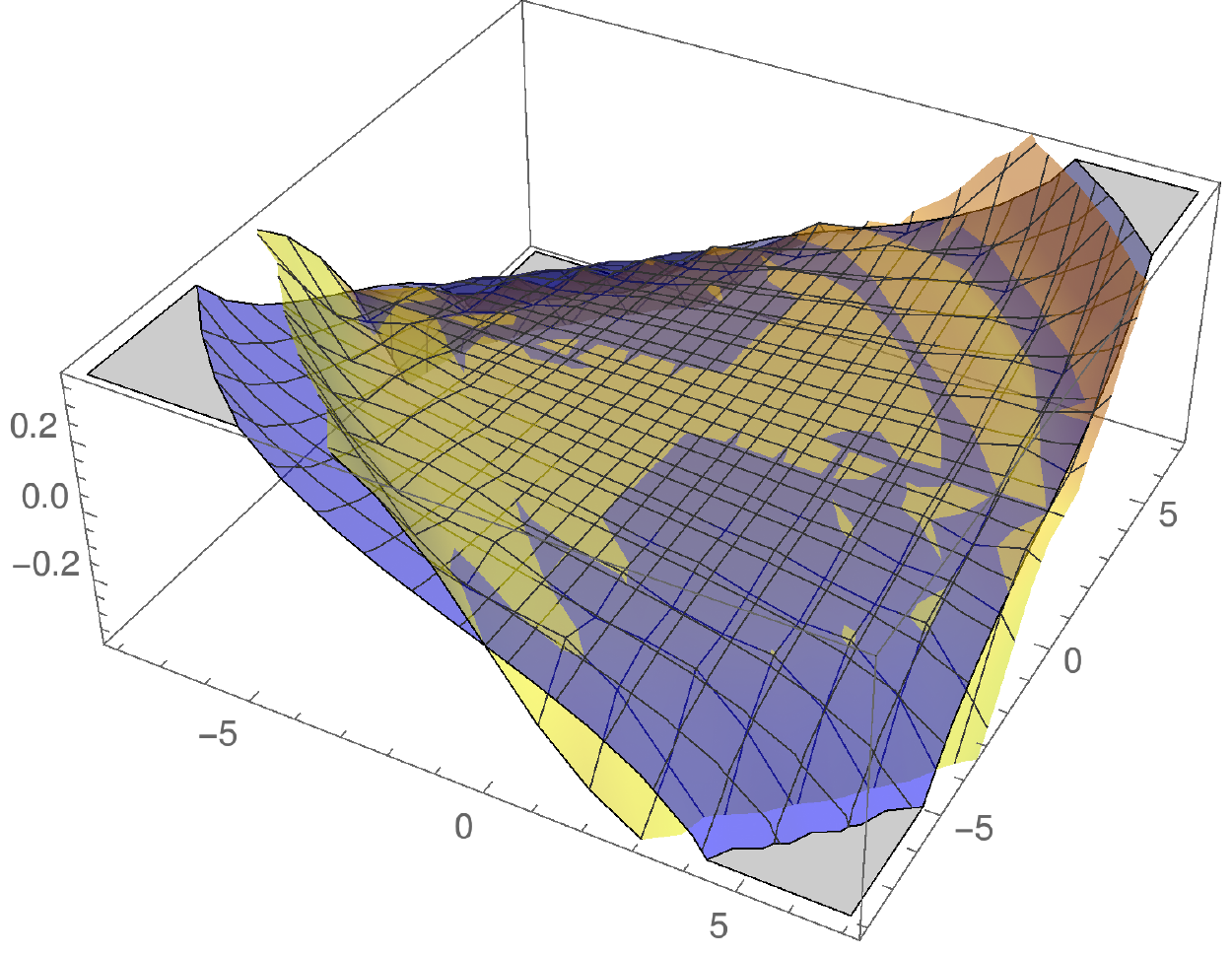} 
\end{center}
\caption{Averaged coefficients of the differential operator $H$ as a function of the position in the $20\times 20$ lattice, with $(0,0)$ being the centre (left: $H_{xx}$ ; right: $2H_{xy}$). The transfer matrix measures (in yellow) are compared to the discrete derivatives of the free energy $F$ calculated from the Bethe ansatz (in blue).}
\label{fig:moyH}
\end{figure}
\noindent each point is averaged with its four neighbours according to the following scheme: $\tfrac{1}{2}(x,y)+\tfrac{1}{8}((x+1,y)+(x-1,y)+(x,y+1)+(x,y-1))$.

The coefficient $H_{yy}$ is equal to $H_{xx}$ because of the symmetry of the isotropic six-vertex model. The coefficient $H_1$ is measured to be of order $10^{-3}$, which is in good agreement with the fact that it should be zero in our description. The coefficients $H_x$ and $H_y$ are small both in the Bethe ansatz and the transfer matrix study (of order $10^{-2}$), but we do not have an agreement as satisfactory as for $H_{xx}$ and $H_{xy}$. The fact that they are of order ten times smaller than the dominant coefficients $H_{xx},H_{yy},H_{xy}$ certainly plays a role, being possibly scrambled by the oscillations or the averaging.

% There, we display the averaged measured coefficient $H_{xx}$ for different sizes, along the diagonal of the lattice. The coefficients $H_{xx,xy}(m_x,m_y)$ calculated from the Bethe ansatz procedure of section \ref{sec:evaluationF} are evaluated on the height function $h_{N=20}$ of the $20\times 20$ square lattice. At $(0,0)$ the magnetizations are zero for all lattice sizes.

We study finite-size effects in Fig. \ref{fig:conv}. We show evidence for the convergence of the coefficients as $N$ increases, by looking at the values on the diagonal of the lattice. In particular the ratio of the measured coefficient over the Bethe ansatz one clearly approaches $1$ quite fast for $c=1/2$ (top right). The value of $H_{xx}$ at the center of the lattice also fits well with the Bethe ansatz calculation. As $c$ increases (or as $\Delta$ decreases) the finite-size effects become stronger and a parity effect becomes visible (bottom right), but the ratio is still close to $1$. This has to do with the fact that the first irrelevant operator becomes less and less irrelevant as $\Delta$ decreases, up to becoming marginal when $\Delta=-1$. In Fig. \ref{fig:rawH} the oscillations happen as well to increase when $\Delta$ decreases, and similar figures would be unreadable.

 \begin{figure}[H]
 \begin{center}
\includegraphics[scale=0.6]{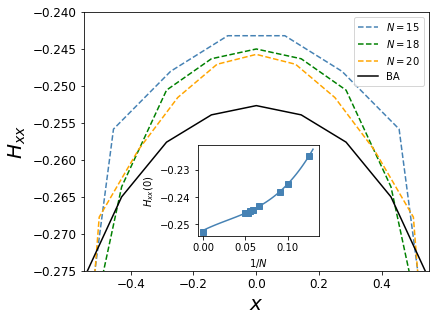}
\includegraphics[scale=0.45]{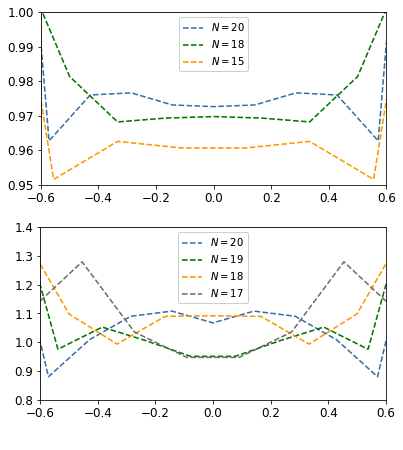}
\end{center}
\caption{Left: averaged measured coefficient $H_{xx}$ on the diagonal of the square lattice $(xN/2,xN/2)$ as a function of the normalized position $x$, for different sizes at $c=1/2$. The black curve is computed with the Bethe ansatz, using the height function of the $20\times 20$ square lattice. Inset: value of $H_{xx}$ at the centre of the lattice $(0,0)$ as a function of $1/N$, together with a polynomial extrapolation and its limit value computed with the Bethe ansatz. Right: same averaged measured coefficient $H_{xx}$ as a function of $x$, but divided by the Bethe ansatz computation using the height function of the same lattice size, for $c=1/2$, i.e. $\Delta=0.875$ (top) and $c=19/10$, i.e. $\Delta\approx -0.8$ (bottom).}
\label{fig:conv}
\end{figure}

To conclude this section, we have provided direct numerical evidence, from finite-size lattice calculations, that correlation functions of the height field are indeed captured by the action (\ref{eq:action0}), or equivalently (\ref{eq:action}), in the the thermodynamic limit.

\newpage

\section{Position dependent GFF coupling constant in the interacting case}
%{In the interacting case, the coupling constant of the Gaussian Free Field is position-dependent}
\label{sec:K}

In this section we establish our second claim, namely that the coupling constant of the Gaussian Free Field (GFF) is position dependent in the interacting case,
$K=K({\rm x})$. Since we have already established above that the magnetization profile $(m_x,m_y)$ varies with position, it is sufficient to establish
that $K(m_x,m_y)$ is a non-constant function in the interacting case.

%To fix the setting, we begin in Section~\ref{sec:K_free} by establishing that $K=1$ is constant in the free-fermion case $\Delta=0$.
%We next provide in Section~\ref{sec:analytic_K}  a number of analytical results on $K(m_x,m_y)$ and its first two derivatives in the interacting case, at a few selected points $(m_x,m_y) = (0,0)$, $(1,m_y)$, and $(m_x,1)$. These results are sufficient to show that $K$ is indeed non-constant, but they fall short of determining the function $K(m_x,m_y)$ completely. Note however that they predict a change of behaviour as a function of $\Delta$: when $\Delta < -\frac12$ the function $K(m_x,m_y)$, considered as a surface, has a singular point at the origin.

\subsection{Numerical evaluation of the coupling constant $K(m_x,m_y)$}
\label{sec:num_K}

In this section we present a numerical calculation of $K(m_x,m_y)$ obtained with the numerical Bethe ansatz method developed in Sec. \ref{sec:evaluationF}. Fig. \ref{figKk} shows the function $K(m_x,m_y)$ for four different values of $\Delta$.
 \begin{figure}[H]
 \begin{center}
\includegraphics[scale=0.6]{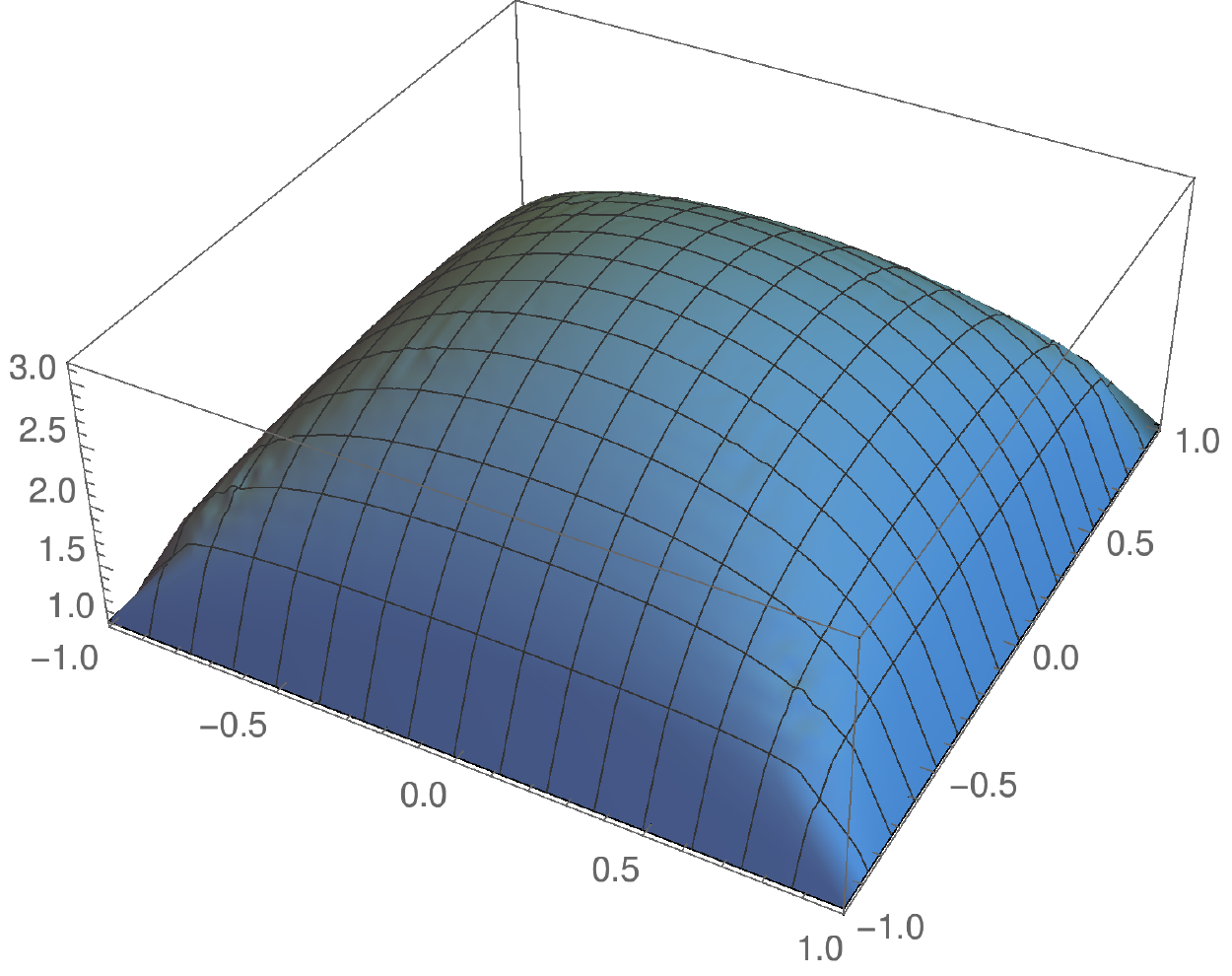} 
\includegraphics[scale=0.6]{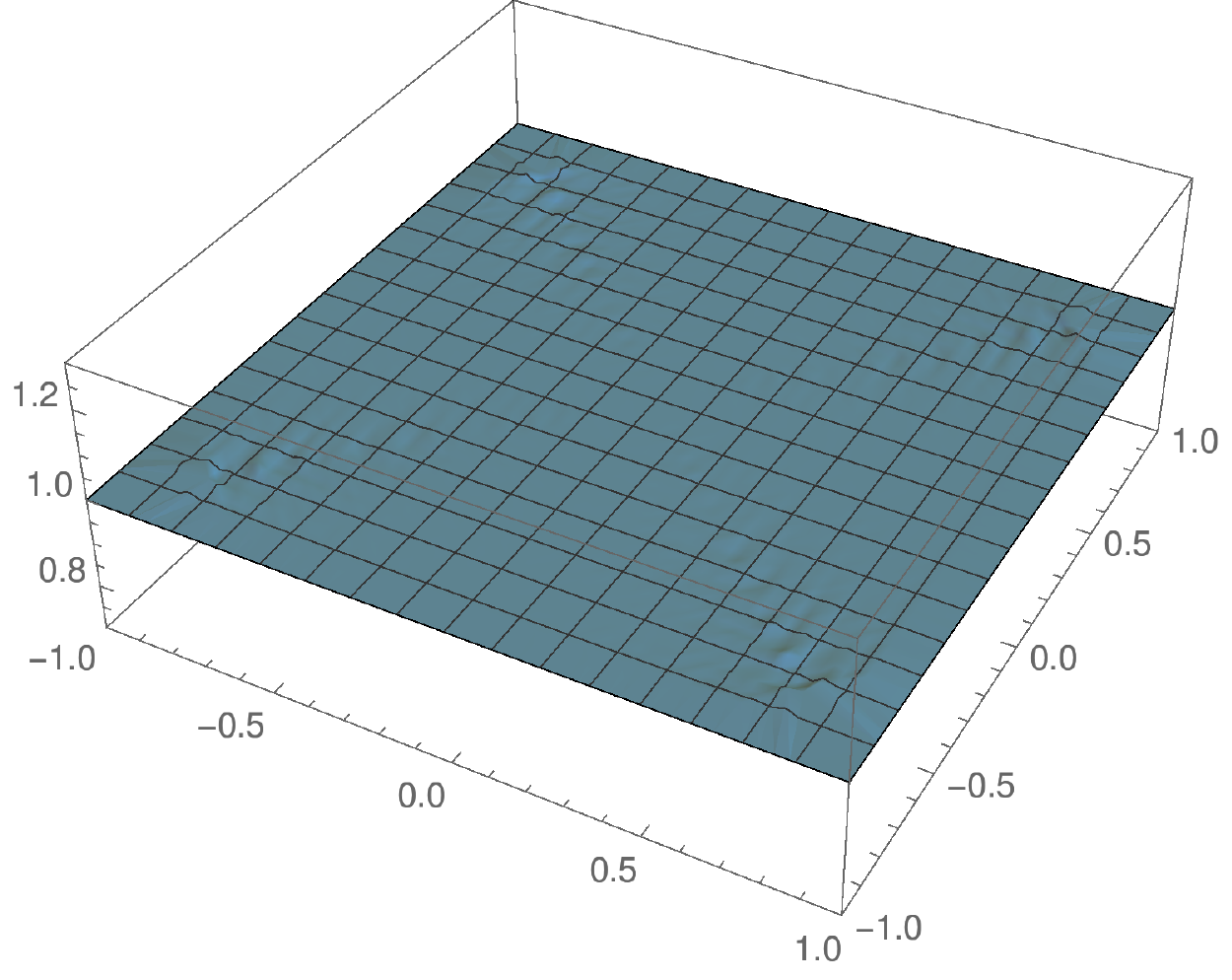} 
\includegraphics[scale=0.6]{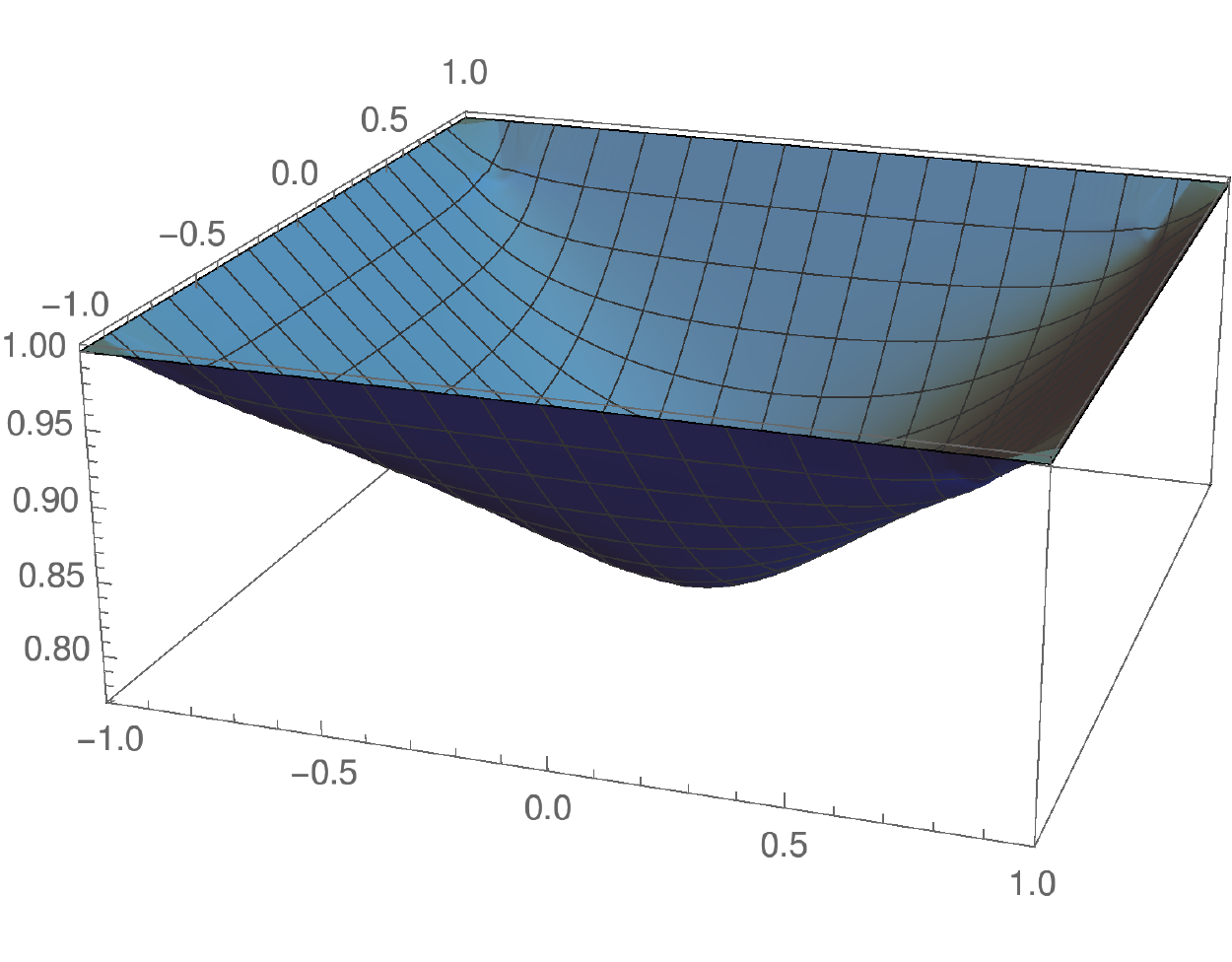} 
\includegraphics[scale=0.6]{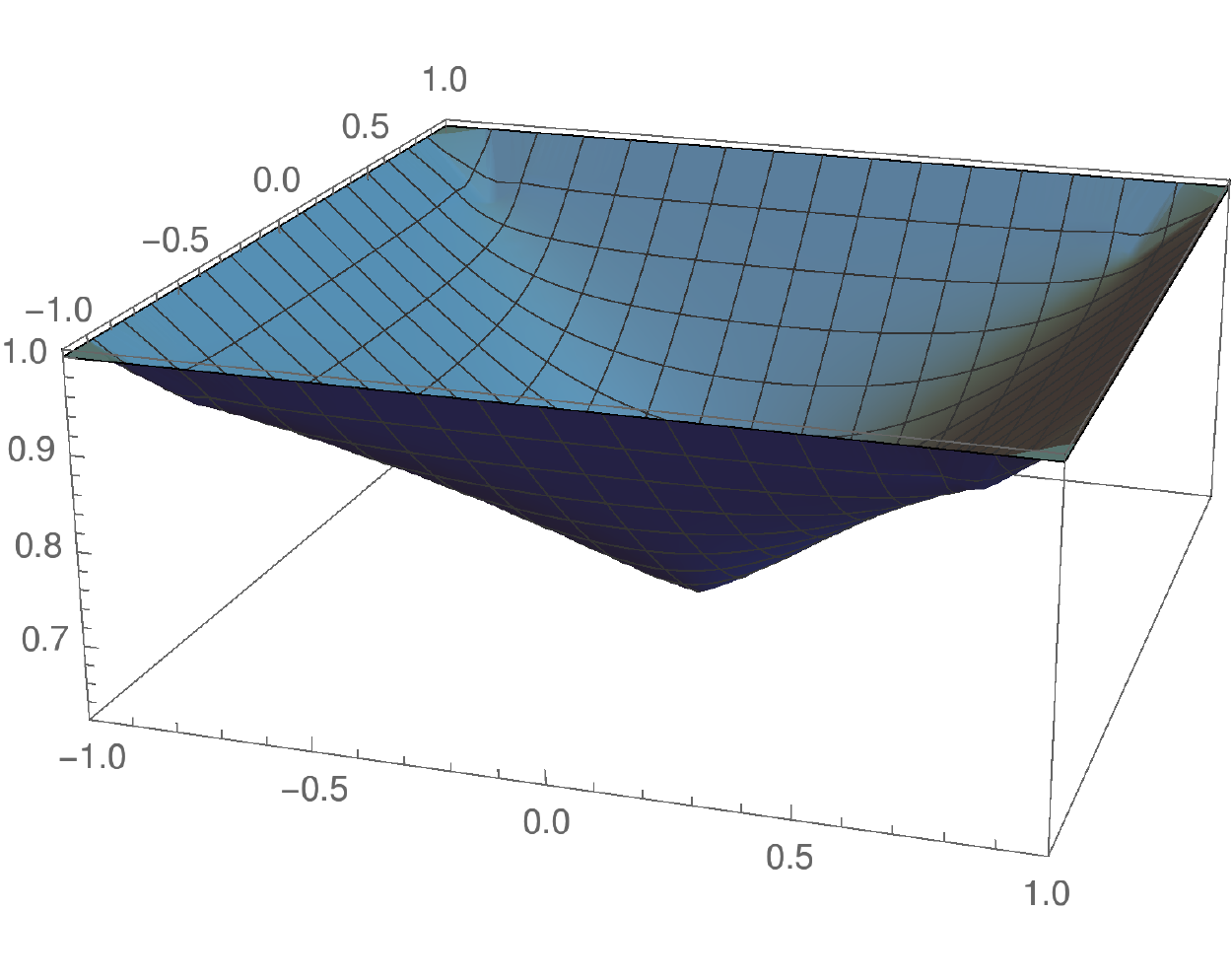} 
\end{center}
\caption{$K$ as a function of $m_x/\pi$ and $m_y/\pi$ for $c=1/2$, i.e. $\Delta=0.875$ (top left), $c=\sqrt{2}$, i.e. $\Delta=0$ (top right), $c=17/10$, i.e. $\Delta= -0.445$ (bottom left) and $c=19/10$, i.e. $\Delta\approx -0.8$ (bottom right).}
 \label{figKk}
\end{figure}
Four qualitatively different cases are plotted: $\Delta>0$ (top left), $\Delta=0$ (top right), $-1/2<\Delta<0$ (bottom left) and $\Delta\leq -1/2$ (bottom right). It is seen that $K$ is concave if $\Delta>0$ and convex if $\Delta<0$, in agreement with analytical results (Eqs.\ \eqref{K0} and \eqref{K1}) that we shall derive below. These numerical results very clearly demonstrate the non-constant nature of $K(m_x,m_y)$. Morevover, they suggest that for $\Delta>-1/2$ the function is smooth (at least twice differentiable), and that for $\Delta\leq -1/2$ the point $(0,0)$ becomes a singular point, a phenomenon that we explore in more detail in subsection \ref{sec:K00_double}.

The small numerical oscillations observed in Fig. \ref{figKk} come from the finite-size estimation of the Hessian of $F$. They decrease as the lattice spacing decreases.

\subsection{Analytic results for $K$ at some special points}
\label{sec:analytic_K}

Despite the fact that for $\Delta\neq 0$, an analytical calculation of $K(m_x,m_y)$ presently seems beyond reach, the value of $K$ and its first two derivatives at special points can be determined analytically. It turns out that this is sufficient to uncover a number of interesting features, and also provides a useful benchmark for the numerical evaluations. We thus analytically study a few features of the function $K(m_x,m_y)$ in this subsection.

\subsubsection{Value of $K(0,0)$ \label{K00}}
To evaluate analytically $F(m_x,m_y)$ one has to know first the free energy $f(m_x,\varphi)$ with given magnetization $m_x$ and parameter $\varphi$. For this we need to know the Bethe root configuration and the Bethe root density in the thermodynamic limit. But as explained in Sec. \ref{sec:roots}, no analytical expressions are known as soon as $\varphi>0$. However, to determine $K(0,0)$ one only needs to know the behaviour of $f(m_x,\varphi)$ for small values of $m_x$ and $\varphi$.

The following fact is known: if $N_x/2-n$ roots $\lambda_i$ are solutions of the following Bethe equations:
\begin{equation}
\left(\dfrac{\sinh(\lambda_i+i\gamma/2)}{\sinh(\lambda_i-i\gamma/2)} \right)^{N_x}=e^{2 i\tilde{\varphi}}\prod_{j\neq i}\dfrac{\sinh(\lambda_i-\lambda_j+i\gamma)}{\sinh(\lambda_i-\lambda_j-i\gamma)}\,,
\end{equation}
with a twist $\tilde{\varphi}$ (without $N_x$), then the free energy per unit area (or vertex) $g_{N_x}$ at size $N_x$ behaves as
\begin{equation}
\label{finitesize}
g_{N_x}=g_\infty-\frac{2\pi}{N_x^2}\left( \frac{1}{12}-\frac{n^2}{2}(1-\gamma/\pi)-\frac{\tilde{\varphi}^2}{2\pi^2}\frac{1}{1-\gamma/\pi}\right)+o(N_x^{-2})\,,
\end{equation}
where $g_\infty$ denotes the corresponding free energy in the thermodynamic limit. Note that this $N_x^{-2}$ finite-size effect actually reads $2\pi (-c/12+h+\bar{h})$ \cite{BloteCardy,Affleck} where $c=1$ is the central charge and $h,\bar{h}$ the conformal weights of the underlying conformal field theory. The twist $\tilde{\varphi}$ appears in the effective central charge that reads $1-6(\tilde{\varphi}/\pi)^2(1-\gamma/\pi)^{-1}$. Here we have $m_x=2 n/N_x$ and $\varphi=i\tilde{\varphi}/N_x$ that are close to zero for large $N_x$. At small but finite magnetization and twist in the thermodynamic limit, we get (with an interchange of limits)
\begin{equation}
f(m_x,\varphi)=f(0,0)+\frac{\pi}{4}(1-\gamma/\pi)m_x^2-\frac{1}{\pi}\frac{1}{1-\gamma/\pi}\varphi^2+o(m_x^2,\varphi^2)\,.
\end{equation}
Using $\partial_\varphi f=m_y$ we get the expansion:
\begin{equation}
F(m_x,m_y)=F(0,0)+\frac{\pi}{4}(1-\gamma/\pi)m_x^2+\frac{\pi}{4}(1-\gamma/\pi)m_y^2+o(m_x^2,m_y^2)\,,
\end{equation}
from which $K(0,0)$ is readily obtained
\begin{equation}
\label{K0}
K(0,0)=\frac{1}{2(1-\gamma/\pi)}\,.
\end{equation}
Remark that this formula is the same as the one for the Luttinger parameter $K$ in the XXZ spin chain at half-filling, see e.g. \cite{Sirker}. In this case in the middle of the lattice the system acts as if the particular boundary conditions imposed did not make any difference. Actually by symmetry, in the middle one has $m_x=m_y=0$, thus $\varphi=0$ and the transfer matrix \eqref{transfer matrix} is nothing but the usual XXZ transfer matrix.

Apart from leading to $K(0,0)$, this computation indicates that if the interchange of limits is possible, to obtain derivatives of this same quantity one would need to know the finite-size corrections
of $g_{N_x}$ to higher order. We have no argument to state that this interchange is possible at all orders, but it works at the order considered in Section~\ref{sec:K00_double} below. We note that anyway the problem of finding all the corrections in $g_{N_x}$ is a notorious unsolved and difficult problem for the interacting case.

\subsubsection{Value of $K(1,m_y)$ and $K(m_x,1)$ \label{secK1}}
The limits where one of the magnetizations goes to $\pm 1$ is of interest since it is observed close to the arctic curve. In our method the two magnetizations are treated very differently. Sending $m_x\to 1$ corresponds to taking very few Bethe roots. In \eqref{be2} it implies that when $N_x\to\infty$ the interaction term between the roots becomes negligible compared to the $\varphi$ term and the source term. Sending $m_y\to 1$ corresponds to taking $\varphi\to\infty$. The Bethe roots then condensate around the point $i\gamma/2$, becoming closer when $N_x\to\infty$. Then the interaction term in \eqref{be2} becomes also negligible since the differences $\lambda_i-\lambda_j$ are small. In both cases the roots are at leading order approximated by the Bethe equations
\begin{equation}
\left(\dfrac{\sinh(\lambda_i+i\gamma/2)}{\sinh(\lambda_i-i\gamma/2)} \right)^{N_x}=e^{2 N_x\varphi}\,,
\end{equation}
which is reminiscent of the free fermion case, except for the $\gamma/2$ instead of $\pi/4$ in the source term. This case becomes exactly solvable and similar calculations to the previous section can be carried out. One finally gets
\begin{equation}
\label{K1}
K(1,m_y)=K(m_x,1)=1\,.
\end{equation}
From these two special points, \eqref{K0} and \eqref{K1}, we already see that $K$ has to vary inside the arctic curve and that the fluctuations can no longer be described by a homogenous Gaussian Free Field.

\subsubsection{Value of $\partial_{m_x}K(1,m_y)$}
The limit $m_x\to 1$ corresponds to taking only few Bethe roots. However one should be careful about the order of the limits: one has to first take the limit $N_x\to\infty$ and then $m_x\to 1$. Writing the Bethe equations \eqref{be2} in logarithmic form for $k$ roots
\begin{equation}
s(\lambda_i)=\frac{I_i}{N_x}+i\frac{\varphi}{\pi}+\frac{1}{N_x}\sum_{j=1}^k r(\lambda_i-\lambda_j)\,,
\end{equation}
with 
\begin{equation}
s(\lambda)=\dfrac{1}{\pi}\arctan \left(\dfrac{\tanh \lambda}{\tan \gamma/2} \right), \quad r(\lambda)=\dfrac{1}{\pi}\arctan \left(\dfrac{\tanh \lambda}{\tan \gamma} \right)\,,
\label{s-r-functions}
\end{equation}
one can determine the Bethe roots $\lambda_i$ perturbatively when the Bethe numbers $I_i/N_x$ and $\varphi$ are assumed to be small. At leading order
\begin{equation}
\lambda_i=\left( \frac{I_i}{N_x}\pi+i\varphi\right)\tan\gamma/2+O((I_i/N_x)^2,\varphi^2)\,.
\end{equation}
This approximation corresponds to the case described in section \ref{secK1}. Then at next-to-leading order
\begin{equation}
\lambda_i=\tan\gamma/2 (\pi I_i/N_x+i\varphi)+\frac{1}{N_xs'(0)}\sum_{j=1}^k r'(0)(\lambda_i-\lambda_j)\,,
\end{equation}
which gives
\begin{equation}
\lambda_i=\left( \pi\frac{I_i}{N_x}\left(1+\left(1-m_x \right)\frac{\tan\gamma/2}{2\tan\gamma}\right)+i\varphi\right)\tan\gamma/2+O((I_i/N_x)^3,\varphi^3)\,.
\end{equation}
This approximation is already excellent in numerical simulations, but two obstacles hinder analytical calculations then: the double limit $N_x\to\infty$ and $m_x\to 1$, and the fact that in the log of \eqref{su2} there is no exponentially dominant term for small $\varphi$. However the following argued conjecture can be made. The next-to-leading order perturbation of the Bethe roots actually corresponds to rescaling the $m_x$ direction by a $m_x$-dependent factor $1+\left(1-m_x \right)\frac{\tan\gamma/2}{2\tan\gamma}$. Since $K$ has the same dimensions as the inverse of the free energy $F$, and since the rescaling in the $m_x$ direction is itself linear in $m_x$, one can expect the behaviour
\begin{equation}
K(m_x,0)=1-\left(1-m_x \right)\frac{\tan\gamma/2}{\tan\gamma}+O((1-m_x)^2)\,.
\end{equation}
Thus we should have
\begin{equation}
\label{Kdpi}
\partial_{m_x}K(1,0)=\frac{\tan\gamma/2}{\tan\gamma}\,.
\end{equation}
Since for $m_x\to 1$, the $\varphi$ that corresponds to a finite $m_y\neq 1,-1$ goes to zero, we conjecture that the same formula applies for all $-1<m_y<1$
\begin{equation}
\label{Kdpi2}
\partial_{m_x}K(1,m_y)=\frac{\tan\gamma/2}{\tan\gamma}\,.
\end{equation}
This computation also suggests that solving perturbatively the Bethe equations at higher-order in $m_x,\varphi$ could permit to compute higher-order derivatives.

In Fig. \ref{fighessian} we compare the analytical formula \eqref{Kdpi} (and also formula \eqref{K20} below) to the numerics. Once again, we need to treat the double limit $N_x\to\infty$ and $m_x\to 1$ carefully: the plots have been obtained by fixing first $m_x=1-1/50$ for the first one and $m_x=1/50$ for the second one, and then taking large $N_x$. The agreement is seen to be excellent.
 \begin{figure}[H]
 \begin{center}
\includegraphics[scale=0.4]{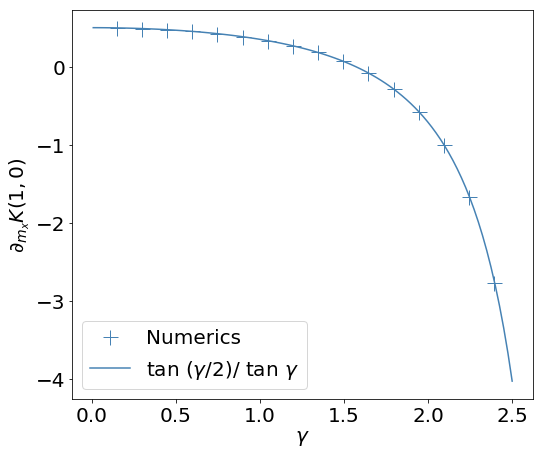} 
\includegraphics[scale=0.4]{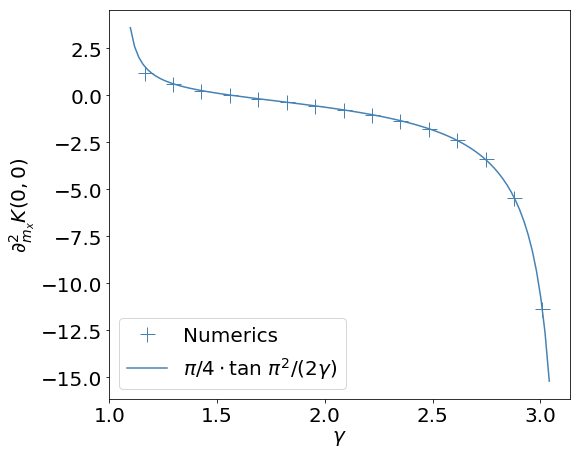} 
\end{center}
\caption{Left: $\partial_{m_x}K(1,0)$ as a function of $\gamma=\arccos -\Delta$, see Eq. (\ref{Kdpi}). Right: $\partial^2_{m_x}K(0,0)$ as a function of $\gamma$, see Eq. (\ref{K20}).}
 \label{fighessian}
\end{figure}

\subsubsection{Value of $\partial^2_{m_x}K(0,0)$}
\label{sec:K00_double}

The reasoning done in section \ref{K00} can be continued in the following way. When $\Delta>-1/2$ the next-order correction to \eqref{finitesize} is of order $O(N_x^{-4})$, and a field theory prediction for the coefficient is given in \cite{Lukyanov}. With an interchange of limits, at small but finite magnetization and twist we get
\begin{equation}
\begin{aligned} \label{resluky}
&f(m_x,\varphi)=f(0,0)+\pi\frac{1-\gamma/\pi}{4}m_x^2-\frac{1}{\pi-\gamma}\varphi^2\\
&-4\pi^2\tan \frac{\pi^2}{2\gamma} \left(\frac{m_x^2}{16}(1-\gamma/\pi)+\frac{\varphi^2}{4\pi^2(1-\gamma/\pi)} \right)^2 \\
&-8\pi^3\lambda_-\left( \frac{(1-\gamma/\pi)^2}{128}m_x^4-\frac{3}{16\pi^2}m_x^2\varphi^2+\frac{1}{8\pi^4(1-\gamma/\pi)^2}\varphi^4 \right)\\
&+o(m_x^4,\varphi^4)\,,
\end{aligned}
\end{equation}
with $\lambda_-$ denoting a coefficient given in \cite{Lukyanov}. When $\Delta<-1/2$ the leading-order correction is of order $O(N_x^{-\delta})$ with $2<\delta<4$ that depends on $\gamma$, and the function is not smooth anymore. We restrict ourselves to the case $\Delta>-1/2$, where the function $f$ is smooth at order $4$ in $m_x,\varphi$.

 One can then perturbatively compute the Legendre transform of this function at order $4$ in $m_x$ and $m_y$, in order to get $K(m_x,m_y)$ at order $2$ in the neighbourhood of $(0,0)$. The (complicated) coefficient $\lambda_-$ turns out to not enter the final result and after some computations we get
\begin{equation}
\label{K20}
\partial^2_{m_x}K(0,0)=\frac{\pi}{4}\tan \frac{\pi^2}{2\gamma} \quad\quad \text{for }\gamma>\frac{\pi}{3}\,, \quad \text{i.e. }\Delta>-\frac{1}{2}\,,
\end{equation}
see also Fig. \ref{fighessian} for a direct comparison with the numerics. \\

Finally, we come back to the singular point at the origin observed in Fig. \ref{figKk} (bottom) for $\Delta \approx -0.8$. We suggest that the singular point at zero is present for all $\Delta\leq -1/2$ (albeit in a degenerate form for $\Delta=-1/2$). Namely, the second derivative of $K$ at zero diverges for all $\Delta\leq -1/2$. For $\Delta<-1/2$, the slope of the singular point at the origin (observed in Fig. \ref{figKk}) seems to go to $0$ as $\Delta\to-1/2$, and somehow hides the transition. Notice that this change of regime is in agreement with the fact that the function \eqref{K20} diverges at $\gamma=\pi/3$. For $\gamma<\pi/3$ another term takes over the $O(m_x^4,\varphi^4)$ term in \eqref{resluky} and the previous derivation is not valid anymore.

To give further evidence for this property, in Fig. \ref{divergence} we show a more precise plot of $K(m_x,0)-K(0,0)$ for values of $\Delta$ around $-1/2$. On the first plot the blue and orange curve appear to be singular at zero: the right and left derivatives seem to remain finite at zero, but are smaller as $\Delta\to -1/2$. This latter fact makes it difficult to identify the singular point without ambiguity. For this reason we plotted as well the quantity $\tfrac{K(m_x,0)-K(0,0)}{|1/2+\Delta|}$, that reveals a change of regime around $\Delta=-1/2$. For $\Delta>-1/2$ the curves are similar to parabolas whereas for $\Delta<-1/2$ all the curves superimpose and show a singular point. This also gives evidence for the fact that the slope behaves as $|1/2+\Delta|$ for $\Delta$ close to $-1/2$.

 \begin{figure}[H]
 \begin{center}
\includegraphics[scale=0.4]{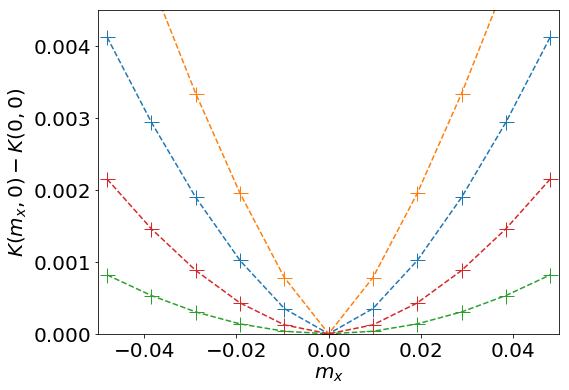} 
\includegraphics[scale=0.4]{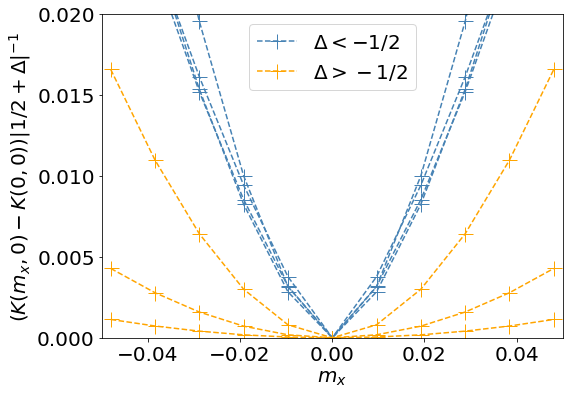} 
\end{center}
\caption{Left: $K(m_x,0)-K(0,0)$ as a function of $m_x$, for $\gamma=\pi/4,\pi/3.5,\pi/3,\pi/2.5$, ie $\Delta\approx-0.7,-0.62,-0.5,-0.3$ (orange, blue, red, green). Right: $\tfrac{K(m_x,0)-K(0,0)}{|1/2+\Delta|}$ as a function of $m_x$, for $\gamma=\pi/4,\pi/3.75,\pi/3.5,\pi/3.25$, $\pi/2.75,\pi/2.5,\pi/2.25$, ie $\Delta\approx -0.7, -0.67, -0.6, -0.57$, $-0.42,-0.3,-0.17$.}
\label{divergence}
\end{figure}

\subsection{$K=1$ in the non-interacting case ($\Delta = 0$)}
\label{sec:K_free}

Finally, we come back to the free-fermion case $\Delta = 0$, and explain why $K=1$ is a constant in that case, contrary to the generic interacting case. The fact that $K=1$ is required in order to be able to fermionize the bosonic degrees of freedom is well-known; for a short explanation of this see for instance Ref. \cite{DubailStephanCalabrese}, or references therein. Here, instead of general arguments based on bosonization/fermionization, we derive the fact that $K=1$ directly from the exact free energy calculated in Sec. \ref{sec:symm_free_energy}.

Recall that the coupling constant $K(m_x,m_y)$ is defined out of  
second derivatives of the free energy $F(m_x,m_y)$ via
\begin{equation}
 \label{K_from_Hess_1} 
K(m_x,m_y)= \left(4 \pi \sqrt{\det \text{Hess } F (m_x,m_y)}\right)^{-1},
\end{equation}
where the Hessian was defined in Eq. (\ref{eq:defhessian}) in the introduction with a non-standard normalization.
We can thus build on the work of Sec. \ref{sec:symm_free_energy}. Since $F$ was shown to be symmetric in its arguments we have: 
\begin{equation}
\partial^2_{m_y} F=-(\partial_{m_y} \varphi)(m_x,m_y), \quad \partial^2_{m_x} F=-(\partial_{m_y} \varphi)(m_y,m_x), \quad \partial_{m_x}\partial_{m_y} F=-(\partial_{m_x} \varphi)(m_x,m_y)\,,
\end{equation}
where we note the flipped arguments in the second expression. In terms of the variables $\alpha$ and $\beta$ introduced in Sec. \ref{sec:symm_free_energy},  the derivatives are
\begin{equation}
\begin{aligned}
\partial_{m_x}\varphi &=\frac{\pi}{4}\dfrac{\alpha\beta}{\sqrt{1-\alpha^2\beta^2}} \,. \\
\partial_{m_y}\varphi &=-\frac{\pi}{4}\sqrt{\dfrac{1-\alpha^2}{1-\beta^2}}\dfrac{1}{\sqrt{1-\alpha^2\beta^2}}\,.
\end{aligned}
\end{equation}
Indeed, the first expression was already given in \eqref{F_cross_derivative}, and the second can be checked by noticing that the
cross derivative matches between the two expressions, $\partial_{m_x} \partial_{m_y} \varphi = \partial_{m_y} \partial_{m_y} \varphi$.
Injecting the two expressions into Eq. \eqref{K_from_Hess_1}, with the normalization of Eq. (\ref{eq:defhessian}), one gets
\begin{equation}
K(m_x,m_y)=1\,.
\end{equation}
Thus, at the free-fermion point, the coupling constant of the GFF is constant (it does not depend on position, contrary to the interacting case), and we thus recover the fact that the fluctuations inside the arctic circle are governed by a standard (i.e. homogeneous) GFF.

\section{Conclusion}
\label{sec:conclusion}

The main goal of this paper was to provide numerical evidence for the two claims stated in the introduction:
(i) the fluctuations of the height field inside the arctic curve in the interacting six-vertex model are Gaussian and described by the Hessian of the free energy $F(m_x,m_y)$, and (ii) the coupling constant $K$ of this Gaussian Free Field varies with position.

To do this, we compared a numerical Bethe ansatz calculation of the Hessian of $F$ to the discretized differential operator that describes the fluctuations obtained from a transfer matrix study in a finite-size lattice. We also provided evidence that Wick's theorem is satisfied in the thermodynamic limit. These two claims imply that the correlation functions of the height-field inside the arctic curve are expressed in terms of the Green's function of a generalized Laplacian, $\Delta = \nabla \cdot \frac{1}{K} \nabla$.\\

To determine numerically the free energy $F(m_x,m_y)$ and then the function $K(m_x,m_y)$ with high precision we used the Bethe ansatz in the XXZ spin chain with an imaginary extensive twist. Tuning the twist imposes the magnetization $m_y$, the other one $m_x$ being imposed by the sector on which the trace is taken. It leads to remarkably symmetric functions $F$ and $K$ in spite of the asymmetric treatment of $m_x$ and $m_y$, and we gave a numerical calculation of the function $K(m_x,m_y)$. This method also permitted us to find analytical expressions for $K$ at some special points. Determining the general expression of $K(m_x,m_y)$ seems very hard (for example the second derivative at zero is related to the higher-order finite-size corrections of the XXZ chain, which is a difficult problem). This Bethe ansatz analysis with extensive imaginary twist may be applied to other integrable models (higher spin, higher rank) and to some more general boundary conditions or more general shapes of the domain. \\

Finally, it would be very interesting to relate the arctic curve phenomenon in the interacting six-vertex model to quantum quench problems, as suggested first by Abanov \cite{abanov2006hydrodynamics}. For spin chains that map to free fermions, there is a clear relation, via a Wick rotation, between arctic circle problems and dynamical models of quantum particles in 1d evolving from a domain-wall state $\left|  \uparrow \uparrow \uparrow \uparrow \uparrow \uparrow  \downarrow \downarrow \downarrow \downarrow \downarrow \downarrow \right>$, see Ref. \cite{AllegraDubailStephanViti} for a detailed discussion. It is tempting to try to extend these results to the interacting case, relying on Wick rotation, especially in view of the tremendous progress that has been accomplished on quenches from inhomogeneous initial states in the past two years \cite{bertini2016transport,castro2016emergent,piroli2017transport,bulchandani2018bethe,doyon2017large,ilievski2017ballistic,mazza2018energy}. In particular, the recent Ref. \cite{collura2018analytic} provided an exact solution of the quench from a domain-wall state in the interacting XXZ chain, and it seems likely that this is related to the arctic circle problem in the six-vertex model. On the other hand, St\'ephan \cite{stephan2017return} has showed that at least in the context of quantum return probabilities, that are related to partition functions of the six-vertex model with domain-wall boundary conditions, the interplay between interactions and the Wick rotation could be very subtle and spoil the nice relation between these problems that exist in the non-interacting case. We leave these exciting and timely questions for future work. \\

{\bf Acknowledgements. } We thank F. Colomo, J.-M. St\'ephan and J. Viti for stimulating discussions and for very useful comments on the manuscript. We also thank H. Saleur for a question that lead to the discussion in the appendix. JD thanks Y. Brun, P. Calabrese, P. Ruggiero, J.-M. St\'ephan, and J. Viti for joint work on very closely related topics. 

JD acknowledges financial support from the CNRS-Mission Interdisciplinaire through the D\'efi Infiniti ``MUSIQ''.

\appendix

\section{What exactly is the minimum around which we are expanding?}
\label{app:propermin}

Here we discuss a technical point which we overlooked in the introduction, when we arrived at the action (\ref{eq:action0}) ---or equivalently (\ref{eq:action})---. We use the same notations as in the introduction: $a_0$ is the lattice spacing, and $h$ is the height field on a square of size $L \times L$. 

The basic principle that ultimately leads to the action (\ref{eq:action0}) is locality. Indeed, the six-vertex model is defined in terms of local degrees of freedom and local Boltzmann weights, so the total free energy of a configuration $(m_x,m_y)$ must take the form of an integral over space, 
\begin{equation}
	\label{eq:freeapp}
	F_{\rm tot} \, = \, \int dx dy  F ( m_x , m_y ,  \partial_x m_x , \partial_x m_y ,  \partial_y m_x,  \partial_y m_y, \partial^2_x m_x,  \partial_x^2 m_y ,  \partial_x \partial_y m_x, \dots ),
\end{equation}
where the integrand depends on the components of the magnetization configuration $m_x(x,y)$, $m_y(x,y)$ and their derivatives. Here we explicitly indicate the dependence on all derivatives, while in the introduction we acted as if $F(m_x, m_y)$ did not depend on the derivatives. This is because  the derivatives $\partial_x^p \partial_y^q m_x$ and $\partial_x^p \partial_y^q m_y$ should be of order $O(1/L^{p+q})$ and therefore should go to zero when $L \rightarrow \infty$. Therefore, the function $F(m_x, m_y)$ that appeared in the introduction is, in fact, the function $F ( m_x , m_y ,  \partial_x m_x , \partial_x m_y ,  \partial_y m_x,  \partial_y m_y, \partial^2_x m_x,  \partial_x^2 m_y ,  \partial_x \partial_y m_x, \dots )$ shown here, evaluated at $\partial_x m_x = \partial_x m_y =  \partial_y m_y, \partial^2_x m_x =  \dots =0$. \\

There is, however, a subtlety. It is not clear that taking the thermodynamic limit $L\rightarrow \infty$ first, and then expanding the free energy functional to second order around its minimum, should lead to the same gaussian action as doing things the other way around ---i.e. expanding first around the minimum in finite size $L$, and then taking the limit $L\rightarrow \infty$---. What we really want to do, to arrive at the action (\ref{eq:action0}), is the latter thing: expand first, then take the limit $L\rightarrow \infty$. \\

Let us discuss this in the language of the height field (\ref{eq:height}). The total free energy (\ref{eq:freeapp}) is
\begin{equation}
	F_{\rm tot} \, = \, \int dx dy  F ( \frac{1}{\pi} \partial_x h ,  -\frac{1}{\pi} \partial_y h ,  \frac{1}{\pi} \partial_x^2 h,  -\frac{1}{\pi} \partial_x \partial_y h  ,  -\frac{1}{\pi}\partial_y^2 h,  \frac{1}{\pi}\partial^3_x h,  -\frac{1}{\pi} \partial_x^2 \partial_y h , \dots ).
\end{equation}
Let $h_0$ be the height configuration that minimizes that functional. This height configuration satisfies the Euler-Lagrange
equation
\begin{equation}
	\label{eq:ELlong}
	\left( \partial_x \frac{\partial  F}{\partial (\partial_x h)}  + \partial_y \frac{\partial  F}{\partial (\partial_y h)}  -  \partial^2_x \frac{\partial  F}{\partial (\partial^2_x h)} -  \partial^2_y \frac{\partial  F}{\partial (\partial^2_y h)}  -  \partial_x \partial_y  \frac{\partial  F}{\partial (\partial_x \partial_y h)}  + \dots\right)_{h=h_0}  = 0 .
\end{equation}
By dimensional analysis, $ \frac{\partial F}{\partial (\partial_x^p \partial_y^q h)}$ must scale as $O(a_0^{p+q})$, so that $ \partial^p_x \partial_y^q \frac{\partial F}{\partial (\partial_x^p \partial_y^q h)}$ is of order $O( (a_0/L)^{p+q} )$. Thus, $h_0$ only approximately 
satisfies the Euler-Lagrange equation that would have been obtained assuming that the free energy $F(m_x, m_y)$ is independent of derivatives:
\begin{equation}
	\label{eq:ELshort}
	\left( \partial_x \frac{\partial  F}{\partial (\partial_x h)}  + \partial_y \frac{\partial  F}{\partial (\partial_y h)}  \right)_{h=h_0}  = \, O(a_0/L ) .
\end{equation}
In other words, the true finite-size minimum $h_0$ differs from the minimum $\tilde{h}_0$ that would be obtained by working with the functional $F(m_x, m_y)$ only, that would satisfy Eq. (\ref{eq:ELshort}) exactly:
\begin{equation}
	\label{eq:ELshort_tilde}
	\left( \partial_x \frac{\partial  F}{\partial (\partial_x h)}  + \partial_y \frac{\partial  F}{\partial (\partial_y h)}  \right)_{h=\tilde{h}_0}  = \, 0 .
\end{equation}
The difference between $h_0$ and $\tilde{h}_0$ is of order $O(a_0/L )$. \\

Expanding the total free energy around the two minima $h_0$ and $\tilde{h}_0$ does not lead to the same result; that is why it is important to be careful about this point. Expanding around the true finite-size minimum $h_0$ leads to
\begin{eqnarray}
	\label{eq:expansionF}
	F_{\rm tot}[h_0 + \delta h] &=&  F_{\rm tot}[h_0]  \\
\nonumber	&& +  \int dx dy  \left(   \frac{\partial F}{\partial (\partial_x h)}   \partial_x \delta h  +   \frac{\partial F}{\partial (\partial_y h)} \partial_y \delta h  +   \frac{\partial F}{\partial (\partial_x^2 h)} \partial_x^2 \delta h+  \frac{\partial F}{\partial (\partial_x \partial_y h)} \partial_x \partial_y \delta h+   \frac{\partial F}{\partial ( \partial_y^2 h)}  \partial_y^2 \delta h + \dots \right) \\
\nonumber	&& + \frac{1}{2} \int dx dy  \left(   \frac{\partial^2 F}{\partial (\partial_x h)^2}   (\partial_x \delta h)^2  + 2 \frac{\partial^2 F}{\partial (\partial_x h)\partial (\partial_y h)} (\partial_x  \delta h)( \partial_y  \delta h ) +  \frac{\partial^2 F}{\partial (\partial_y h)^2}   (\partial_y \delta h)^2 +  \dots \right) \\
\nonumber && + \, \dots
\end{eqnarray}
The second line vanishes because of the Euler-Lagrange equation (\ref{eq:ELlong}). Hence arrive at the action (\ref{eq:action0}) in the introduction, by dropping all derivatives that are of order higher than $2$, because these terms are irrelevant in the renormalization group sense.

On the other hand, if one expands around the height configuration $\tilde{h}_0$ that satisfies Eq. (\ref{eq:ELshort_tilde}), we get an expansion identical to (\ref{eq:expansionF}),
%\begin{eqnarray}
%	F_{\rm tot}[\tilde{h}_0 + \delta h] &=&  F_{\rm tot}[\tilde{h}_0]  \\
%\nonumber	&& +  \int dx dy  \left(   \frac{\partial F}{\partial (\partial_x h)}   \partial_x \delta h  +   \frac{\partial F}{\partial (\partial_y h)} \partial_y \delta h  +   \frac{\partial F}{\partial (\partial_x^2 h)} \partial_x^2 \delta h+  \frac{\partial F}{\partial (\partial_x \partial_y h)} \partial_x \partial_y \delta h+   \frac{\partial F}{\partial ( \partial_y^2 h)}  \partial_y^2 \delta h + \dots \right) \\
%\nonumber	&& + \frac{1}{2} \int dx dy  \left(   \frac{\partial^2 F}{\partial (\partial_x h)^2}   (\partial_x \delta h)^2  + 2 \frac{\partial^2 F}{\partial (\partial_x h)\partial (\partial_y h)} (\partial_x  \delta h)( \partial_y  \delta h ) +  \frac{\partial^2 F}{\partial (\partial_y h)^2}   (\partial_y \delta h)^2 +  \dots \right) \\
%\nonumber && + \, \dots
%\end{eqnarray}
but this time the second line does not vanish; only the sum of the first two terms in the second line vanish thanks to Eq. (\ref{eq:ELshort_tilde}). Dropping all the terms with derivatives of order higher than $2$, we arrive at an effective action for the fluctuations that includes terms that are linear in $h$,
\begin{eqnarray}
	&& F_{\rm tot}[\tilde{h}_0 + \delta h] - F_{\rm tot}[\tilde{h}_0]   \\
 \nonumber	&& \qquad =   \int dx dy  \left(   \frac{\partial F}{\partial (\partial_x^2 h)} \partial_x^2 \delta h+  \frac{\partial F}{\partial (\partial_x \partial_y h)} \partial_x \partial_y \delta h+   \frac{\partial F}{\partial ( \partial_y^2 h)}  \partial_y^2 \delta h  \right. \\
\nonumber && \qquad\qquad\qquad \qquad \left. + \frac{1}{2} \frac{\partial^2 F}{\partial (\partial_x h)^2}   (\partial_x \delta h)^2  +  \frac{\partial^2 F}{\partial (\partial_x h)\partial (\partial_y h)} (\partial_x  \delta h)( \partial_y  \delta h ) + \frac{1}{2} \frac{\partial^2 F}{\partial (\partial_y h)^2}   (\partial_y \delta h)^2 \right)  .
%\nonumber	&& +  \int dx dy F ( \frac{1}{\pi} \partial_x h ,  -\frac{1}{\pi} \partial_y h ,  \frac{1}{\pi} \partial_x^2 h,  -\frac{1}{\pi} \partial_x \partial_y h  ,  -\frac{1}{\pi}\partial_y^2 h,  \frac{1}{\pi}\partial^3_x h,  -\frac{1}{\pi} \partial_x^2 \partial_y h , \dots )
\end{eqnarray}
This action is different from the one considered in this paper, see Eqs. (\ref{eq:action0})-(\ref{eq:action}). Thus, it is important to be cautious when one defines the minimum $h_0$.

\bibliography{igff_6v_5}

\begin{thebibliography}{10}

\bibitem{korepin1982calculation}
V.~E. Korepin, ``Calculation of norms of {B}ethe wave functions,'' {\em
  Communications in Mathematical Physics}, vol.~86, no.~3, pp.~391--418, 1982.

\bibitem{AGIzergin}
A.~G. Izergin, ``Partition function of the six-vertex model in the finite
  volume,'' {\em Sov. Phys. Dokl.}, vol.~32, p.~878, 1987.

\bibitem{izergin1992determinant}
A.~G. Izergin, D.~A. Coker, and V.~E. Korepin, ``Determinant formula for the
  six-vertex model,'' {\em Journal of Physics A: Mathematical and General},
  vol.~25, no.~16, p.~4315, 1992.

\bibitem{KorepinZinnJustin}
V.~Korepin and P.~Zinn-Justin, ``Thermodynamic limit of the six-vertex model
  with domain wall boundary conditions,'' {\em J. Phys. A}, vol.~33, 40,
  p.~7053, 2000.

\bibitem{zinn2000six}
P.~Zinn-Justin, ``Six-vertex model with domain wall boundary conditions and
  one-matrix model,'' {\em Physical Review E}, vol.~62, no.~3, p.~3411, 2000.

\bibitem{zinn2002influence}
P.~Zinn-Justin, ``The influence of boundary conditions in the six-vertex
  model,'' {\em arXiv preprint cond-mat/0205192}, 2002.

\bibitem{bogoliubov2002boundary}
N.~Bogoliubov, A.~Pronko, and M.~Zvonarev, ``Boundary correlation functions of
  the six-vertex model,'' {\em Journal of Physics A: Mathematical and General},
  vol.~35, no.~27, p.~5525, 2002.

\bibitem{palamarchuk20106}
K.~Palamarchuk and N.~Reshetikhin, ``The 6-vertex model with fixed boundary
  conditions,'' {\em arXiv preprint arXiv:1010.5011}, 2010.

\bibitem{galleas2010functional}
W.~Galleas, ``Functional relations for the six-vertex model with domain wall
  boundary conditions,'' {\em Journal of Statistical Mechanics: Theory and
  Experiment}, vol.~2010, no.~06, p.~P06008, 2010.

\bibitem{cugliandolo2015six}
L.~F. Cugliandolo, G.~Gonnella, and A.~Pelizzola, ``Six--vertex model with
  domain wall boundary conditions in the {B}ethe--{P}eierls approximation,''
  {\em Journal of Statistical Mechanics: Theory and Experiment}, vol.~2015,
  no.~6, p.~P06008, 2015.

\bibitem{lyberg2018phase}
I.~Lyberg, V.~Korepin, G.~Ribeiro, and J.~Viti, ``Phase separation in the
  six-vertex model with a variety of boundary conditions,'' {\em Journal of
  Mathematical Physics}, vol.~59, no.~5, p.~053301, 2018.

\bibitem{keesman2017numerical}
R.~Keesman and J.~Lamers, ``Numerical study of the {F} model with domain-wall
  boundaries,'' {\em Physical Review E}, vol.~95, no.~5, p.~052117, 2017.

\bibitem{lyberg2017density}
I.~Lyberg, V.~Korepin, and J.~Viti, ``The density profile of the six vertex
  model with domain wall boundary conditions,'' {\em Journal of Statistical
  Mechanics: Theory and Experiment}, vol.~2017, no.~5, p.~053103, 2017.

\bibitem{vershik1977asymptotics}
A.~M. Vershik and S.~V. Kerov, ``Asymptotics of {P}lancherel measure of
  symmetrical group and limit form of {Y}oung tables,'' {\em Doklady Akademii
  Nauk SSSR}, vol.~233, no.~6, pp.~1024--1027, 1977.

\bibitem{logan1982variational}
B.~F. Logan and L.~A. Shepp, ``A variational problem for random {Y}oung
  tableaux,'' in {\em Young Tableaux in Combinatorics, Invariant Theory, and
  Algebra}, pp.~63--79, Elsevier, 1982.

\bibitem{rottman1984statistical}
C.~Rottman and M.~Wortis, ``Statistical mechanics of equilibrium crystal
  shapes: Interfacial phase diagrams and phase transitions,'' {\em Physics
  Reports}, vol.~103, no.~1-4, pp.~59--79, 1984.

\bibitem{nienhuis1984b}
B.~Nienhuis, H.~Hilhorst, and H.~Bl{\"o}te, ``Triangular {SOS} models and
  cubic-crystal shapes,'' {\em J. Phys. A}, vol.~17, p.~3559, 1984.

\bibitem{kenyon2007limit}
R.~Kenyon and A.~Okounkov, ``Limit shapes and the complex {B}urgers equation,''
  {\em Acta mathematica}, vol.~199, no.~2, pp.~263--302, 2007.

\bibitem{colomo2010limit}
F.~Colomo and A.~Pronko, ``The limit shape of large alternating sign
  matrices,'' {\em SIAM Journal on Discrete Mathematics}, vol.~24, no.~4,
  pp.~1558--1571, 2010.

\bibitem{reshetikhin2016limit}
N.~Reshetikhin and A.~Sridhar, ``Limit shapes of the stochastic six vertex
  model,'' {\em arXiv preprint arXiv:1609.01756}, 2016.

\bibitem{kenyon2009lectures}
R.~Kenyon, ``Lectures on dimers,'' {\em arXiv preprint arXiv:0910.3129}, 2009.

\bibitem{reshetikhin2010lectures}
N.~Reshetikhin, ``Lectures on the integrability of the six-vertex model,'' {\em
  Exact methods in low-dimensional statistical physics and quantum computing},
  pp.~197--266, 2010.

\bibitem{jockusch1998random}
W.~Jockusch, J.~Propp, and P.~Shor, ``Random domino tilings and the arctic
  circle theorem,'' {\em arXiv preprint math/9801068}, 1998.

\bibitem{ColomoPronko}
F.~Colomo and A.~Pronko, ``The arctic curve of the domain-wall six-vertex
  model,'' {\em J. Stat. Phys.}, vol.~138, p.~662, 2010.

\bibitem{colomo2010arctic}
F.~Colomo, A.~Pronko, and P.~Zinn-Justin, ``The arctic curve of the domain wall
  six-vertex model in its antiferroelectric regime,'' {\em Journal of
  Statistical Mechanics: Theory and Experiment}, vol.~2010, no.~03, p.~L03002,
  2010.

\bibitem{colomo2011algebraic}
F.~Colomo, V.~Noferini, and A.~Pronko, ``Algebraic arctic curves in the
  domain-wall six-vertex model,'' {\em Journal of Physics A: Mathematical and
  Theoretical}, vol.~44, no.~19, p.~195201, 2011.

\bibitem{colomo2015thermodynamics}
F.~Colomo and A.~G. Pronko, ``Thermodynamics of the six-vertex model in an
  {L}-shaped domain,'' {\em Communications in Mathematical Physics}, vol.~339,
  no.~2, pp.~699--728, 2015.

\bibitem{ColomoSportiello}
F.~Colomo and A.~Sportiello, ``Arctic curves of the six-vertex model on generic
  domains: the {T}angent {M}ethod,'' {\em J. Stat. Phys.}, vol.~164, p.~1488,
  2016.

\bibitem{reshetikhin2017integrability}
N.~Reshetikhin and A.~Sridhar, ``Integrability of limit shapes of the six
  vertex model,'' {\em Communications in Mathematical Physics}, vol.~356,
  no.~2, pp.~535--565, 2017.

\bibitem{di2018arctic}
P.~Di~Francesco and M.~F. Lapa, ``Arctic curves in path models from the tangent
  method,'' {\em Journal of Physics A: Mathematical and Theoretical}, vol.~51,
  no.~15, p.~155202, 2018.

\bibitem{di2018arctic2}
P.~Di~Francesco and E.~Guitter, ``Arctic curves for paths with arbitrary
  starting points: a tangent method approach,'' {\em arXiv preprint
  arXiv:1803.11463}, 2018.

\bibitem{brun2017inhomogeneous}
Y.~Brun and J.~Dubail, ``The {I}nhomogeneous {G}aussian {F}ree {F}ield, with
  application to ground state correlations of trapped 1d {B}ose gases,'' {\em
  arXiv preprint arXiv:1712.05262}, 2017.

\bibitem{noh1996finite}
J.~D. Noh and D.~Kim, ``Finite-size scaling and the toroidal partition function
  of the critical asymmetric six-vertex model,'' {\em Physical Review E},
  vol.~53, no.~4, p.~3225, 1996.

\bibitem{petrov2015asymptotics}
L.~Petrov {\em et~al.}, ``Asymptotics of uniformly random lozenge tilings of
  polygons. {G}aussian free field,'' {\em The Annals of Probability}, vol.~43,
  no.~1, pp.~1--43, 2015.

\bibitem{RKenyon}
R.~Kenyon, ``Conformal invariance of domino tiling,'' {\em Ann. Probab.},
  vol.~29, p.~1128, 2001.

\bibitem{2015arXiv151008248D}
M.~{Duits}, ``{On global fluctuations for non-colliding processes},'' {\em
  ArXiv e-prints}, Oct. 2015.

\bibitem{bufetov2016fluctuations}
A.~Bufetov and V.~Gorin, ``Fluctuations of particle systems determined by
  {S}chur generating functions,'' {\em arXiv preprint arXiv:1604.01110}, 2016.

\bibitem{gorin2017bulk}
V.~Gorin, ``Bulk universality for random lozenge tilings near straight
  boundaries and for tensor products,'' {\em Communications in Mathematical
  Physics}, vol.~354, no.~1, pp.~317--344, 2017.

\bibitem{sheffield2007gaussian}
S.~Sheffield, ``Gaussian free fields for mathematicians,'' {\em Probability
  theory and related fields}, vol.~139, no.~3-4, pp.~521--541, 2007.

\bibitem{AllegraDubailStephanViti}
N.~Allegra, J.~Dubail, J.-M. St\'ephan, and J.~Viti, ``Inhomogeneous field
  theory inside the arctic circle,'' {\em J. Stat. Mech.}, p.~053108, 2016.

\bibitem{dubail2017conformal}
J.~Dubail, J.-M. St{\'e}phan, J.~Viti, and P.~Calabrese, ``Conformal field
  theory for inhomogeneous one-dimensional quantum systems: the example of
  non-interacting {F}ermi gases,'' {\em SciPost Physics}, vol.~2, no.~1,
  p.~002, 2017.

\bibitem{DubailStephanCalabrese}
J.~Dubail, J.-M. St\'ephan, and P.~Calabrese, ``Emergence of curved light-cones
  in a class of inhomogeneous {L}uttinger liquids,'' {\em SciPost Physics},
  vol.~3, p.~019, 2017.

\bibitem{alet1}
F.~Alet, J.~L. Jacobsen, G.~Misguich, V.~Pasquier, F.~Mila, and M.~Troyer,
  ``Interacting classical dimers on the square lattice,'' {\em Phys. Rev.
  Lett.}, vol.~94, p.~235702, 2005.

\bibitem{alet2}
F.~Alet, Y.~Ikhlef, J.~L. Jacobsen, G.~Misguich, and V.~Pasquier, ``Classical
  dimers with aligning interactions on the square lattice,'' {\em Phys. Rev.
  E}, vol.~74, p.~041124, 2006.

\bibitem{Toninelli1}
A.~Giuliani, V.~Mastropietro, and F.~Toninelli, ``Height fluctuations in
  non-integrable classical dimers,'' {\em EuroPhysics Letters}, vol.~109,
  p.~60004, 2015.

\bibitem{Toninelli2}
A.~Giuliani, V.~Mastropietro, and F.~Toninelli, ``Height fluctuations in
  interacting dimers,'' {\em Annales de l'Institut Henri Poincar\'e -
  Probabilit\'es et Statistiques}, vol.~53, p.~98, 2017.

\bibitem{Toninelli3}
A.~Giuliani, V.~Mastropietro, and F.~Toninelli, ``Haldane relation for
  interacting dimers,'' {\em J. Stat. Mech.}, p.~034002, 2017.

\bibitem{AlcarazBarberBatchelor2}
F.~C. Alcaraz, M.~N. Barber, and M.~T. Batchelor, ``Conformal invariance, the
  {XXZ} chain and the operator content of two-dimensional critical systems,''
  {\em Annals of Physics}, vol.~182, p.~280, 1988.

\bibitem{AlcarazBarberBatchelor}
F.~C. Alcaraz, M.~N. Barber, and M.~T. Batchelor, ``Conformal invariance and
  the spectrum of the {XXZ} chain,'' {\em Phys. Rev. Lett.}, vol.~58, no.~8,
  p.~771, 1987.

\bibitem{BloteCardy}
H.~W.~J. Bl{\"o}te, J.~L. Cardy, and M.~P. Nightingale, ``Conformal invariance,
  the central charge, and universal finite-size amplitudes at criticality,''
  {\em Phys. Rev. Lett.}, vol.~56, no.~7, p.~742, 1986.

\bibitem{Affleck}
I.~Affleck, ``Universal term in the free energy at a critical point and the
  conformal anomaly,'' {\em Phys. Rev. Lett.}, vol.~56, no.~7, p.~746, 1986.

\bibitem{Sirker}
J.~Sirker, ``The {L}uttinger liquid and integrable models,'' {\em Int. J. Mod.
  Phys. B}, vol.~26, p.~1244009, 2012.

\bibitem{Lukyanov}
S.~Lukyanov, ``Low energy effective hamiltonian for the {XXZ} spin chain,''
  {\em Nucl.Phys. B}, vol.~522, p.~533, 1998.

\bibitem{abanov2006hydrodynamics}
A.~G. Abanov, ``Hydrodynamics of correlated systems,'' in {\em Applications of
  Random Matrices in Physics}, pp.~139--161, Springer, 2006.

\bibitem{bertini2016transport}
B.~Bertini, M.~Collura, J.~De~Nardis, and M.~Fagotti, ``Transport in
  out-of-equilibrium {XXZ} chains: Exact profiles of charges and currents,''
  {\em Physical review letters}, vol.~117, no.~20, p.~207201, 2016.

\bibitem{castro2016emergent}
O.~A. Castro-Alvaredo, B.~Doyon, and T.~Yoshimura, ``Emergent hydrodynamics in
  integrable quantum systems out of equilibrium,'' {\em Physical Review X},
  vol.~6, no.~4, p.~041065, 2016.

\bibitem{piroli2017transport}
L.~Piroli, J.~De~Nardis, M.~Collura, B.~Bertini, and M.~Fagotti, ``Transport in
  out-of-equilibrium {XXZ} chains: Nonballistic behavior and correlation
  functions,'' {\em Physical Review B}, vol.~96, no.~11, p.~115124, 2017.

\bibitem{bulchandani2018bethe}
V.~B. Bulchandani, R.~Vasseur, C.~Karrasch, and J.~E. Moore,
  ``Bethe-{B}oltzmann hydrodynamics and spin transport in the {XXZ} chain,''
  {\em Physical Review B}, vol.~97, no.~4, p.~045407, 2018.

\bibitem{doyon2017large}
B.~Doyon, J.~Dubail, R.~Konik, and T.~Yoshimura, ``Large-scale description of
  interacting one-dimensional {B}ose gases: generalized hydrodynamics
  supersedes conventional hydrodynamics,'' {\em Physical review letters},
  vol.~119, no.~19, p.~195301, 2017.

\bibitem{ilievski2017ballistic}
E.~Ilievski and J.~De~Nardis, ``Ballistic transport in the one-dimensional
  {H}ubbard model: the hydrodynamic approach,'' {\em Physical Review B},
  vol.~96, no.~8, p.~081118, 2017.

\bibitem{mazza2018energy}
L.~Mazza, J.~Viti, M.~Carrega, D.~Rossini, and A.~De~Luca, ``Energy transport
  in an integrable parafermionic chain via generalized hydrodynamics,'' {\em
  arXiv preprint arXiv:1804.04476}, 2018.

\bibitem{collura2018analytic}
M.~Collura, A.~De~Luca, and J.~Viti, ``Analytic solution of the domain-wall
  nonequilibrium stationary state,'' {\em Physical Review B}, vol.~97, no.~8,
  p.~081111, 2018.

\bibitem{stephan2017return}
J.-M. St{\'e}phan, ``Return probability after a quench from a domain wall
  initial state in the spin-1/2 {XXZ} chain,'' {\em Journal of Statistical
  Mechanics: Theory and Experiment}, vol.~2017, no.~10, p.~103108, 2017.

\end{thebibliography}
\bibliographystyle{ieeetr}
\end{document}